\documentclass[11pt, draftclsnofoot, onecolumn]{IEEEtran}

\usepackage[noadjust]{cite}
\usepackage{cite}
\usepackage{graphicx,amsmath,epsfig,times}
\usepackage{caption,subcaption}
\usepackage{amstext,amssymb,latexsym}
\usepackage{color}
\usepackage{enumitem}
\usepackage{bbm}
\usepackage{mathrsfs}
\usepackage{xspace}
\usepackage{bbm}
\catcode`@=11
\chardef\mathlig@atcode\count255

\def\actively#1#2{\begingroup\uccode`\~=`#2\relax\uppercase{\endgroup#1~}}
\def\mathlig@gobble{\afterassignment\mathlig@next@cmd\let\mathlig@next= }
\def\mathlig@delim{\mathlig@delim}
\def\mathlig@defcs#1{\expandafter\def\csname#1\endcsname}
\def\mathlig@let@cs#1#2{\expandafter\let\expandafter#1\csname#2\endcsname}
\def\mathlig@appendcs#1#2{\expandafter\edef\csname#1\endcsname{\csname#1\endcsname#2}}

\def\mathlig#1#2{\mathlig@checklig#1\mathlig@end\mathlig@defcs{mathlig@back@#1}{#2}\ignorespaces}

\def\mathlig@checklig#1#2\mathlig@end{%
 \expandafter\ifx\csname mathlig@forw@#1\endcsname\relax
 \expandafter\mathchardef\csname mathlig@back@#1\endcsname=\mathcode`#1%
 \mathcode`#1"8000\actively\def#1{\csname mathlig@look@#1\endcsname}%
 \mathlig@dolig#1\mathlig@delim
\fi
\mathlig@checksuffix#1#2\mathlig@end
}

\def\mathlig@checksuffix#1#2\mathlig@end{%
\ifx\mathlig@delim#2\mathlig@delim\relax\else\mathlig@checksuffix@{#1}#2\mathlig@end\fi
}
\def\mathlig@checksuffix@#1#2#3\mathlig@end{%
\expandafter\ifx\csname mathlig@forw@#1#2\endcsname\relax\mathlig@dosuffix{#1}{#2}\fi
\mathlig@checksuffix{#1#2}#3\mathlig@end
}

\def\mathlig@dosuffix#1#2{%
\mathlig@appendcs{mathlig@toks@#1}{#2}%
\mathlig@dolig{#1}{#2}\mathlig@delim
}

\def\mathlig@dolig#1#2\mathlig@delim{%
 \mathlig@defcs{mathlig@look@#1#2}{%
 \mathlig@let@cs\mathlig@next{mathlig@forw@#1#2}\futurelet\mathlig@next@tok\mathlig@next}%
 \mathlig@defcs{mathlig@forw@#1#2}{%
  \mathlig@let@cs\mathlig@next{mathlig@back@#1#2}%
  \mathlig@let@cs\checker{mathlig@chck@#1#2}%
  \mathlig@let@cs\mathligtoks{mathlig@toks@#1#2}%
  \expandafter\ifx\expandafter\mathlig@delim\mathligtoks\mathlig@delim\relax\else
  \expandafter\checker\mathligtoks\mathlig@delim\fi
  \mathlig@next
 }%
 \mathlig@defcs{mathlig@toks@#1#2}{}%
 \mathlig@defcs{mathlig@chck@#1#2}##1##2\mathlig@delim{%
  \ifx\mathlig@next@tok##1%
   \mathlig@let@cs\mathlig@next@cmd{mathlig@look@#1#2##1}\let\mathlig@next\mathlig@gobble
  \fi 
  \ifx\mathlig@delim##2\mathlig@delim\relax\else
   \csname mathlig@chck@#1#2\endcsname##2\mathlig@delim
  \fi
 }%
 \ifx\mathlig@delim#2\mathlig@delim\else
  \mathlig@defcs{mathlig@back@#1#2}{\csname mathlig@back@#1\endcsname #2}%
 \fi
}%

\catcode`@\mathlig@atcode

\newcommand{\muspace}{\mspace{1mu}}

\DeclareRobustCommand{\scond}{\mathchoice{\muspace\vert\muspace}{\vert}{\vert}{\vert}}
\mathlig{|}{\scond}

\DeclareRobustCommand{\discint}{\mathchoice{\mspace{-1.5mu}:\mspace{-1.5mu}}{\mspace{-1.5mu}:\mspace{-1.5mu}}{:}{:}}
\mathlig{::}{\discint}
\newcommand{\suchthat}{\colon}

\newcommand{\Ac}{\mathcal{A}}

\newcommand{\Cc}{\mathcal{C}}
\newcommand{\cc}{\text{\footnotesize $\mathcal{C}$} }

\newcommand{\Ec}{\mathcal{E}}

\newcommand{\Lc}{\mathcal{L}}

\newcommand{\Sc}{\mathcal{S}}

\newcommand{\Uc}{\mspace{1.5mu}\mathcal{U}}

\newcommand{\Xc}{\mathcal{X}}
\newcommand{\Yc}{\mathcal{Y}}

\newcommand{\Rr}{\mathscr{R}}

\newcommand{\av}{{\bf a}}
\newcommand{\bv}{{\bf b}}

\newcommand{\aep}{{\mathcal{T}_{\epsilon}^{(n)}}}
\newcommand{\aepvar}{{\mathcal{T}_{\epsilon'}^{(n)}}}
\newcommand{\aepvarvar}{{\mathcal{T}_{\epsilon''}^{(n)}}}

\newcommand{\Rh}{{\hat{R}}}

\newcommand{\Wh}{{\hat{W}}}

\newcommand{\wh}{{\hat{w}}}

\newcommand{\ut}{{\tilde{u}}}

\def\a{\alpha}

\def\g{\gamma}

\def\d{\delta}
\def\e{\epsilon}

\def\Th{\Theta}

\DeclareMathOperator\E{\textsf{E}}
\let\P\relax
\DeclareMathOperator\P{\textsf{P}}

\newcommand{\U}{\mathrm{Unif}}

\def\textiid{i.i.d.\@\xspace}
\newcommand\iid{\ifmmode\text{ i.i.d. } \else \textiid \fi}

\newcommand{\Zz}{\mathbb{Z}}
\newcommand{\Real}{\mathbb{R}}

\def\clap#1{\hbox to 0pt{\hss#1\hss}}
\def\mathclap{\mathpalette\mathclapinternal}
\def\mathclapinternal#1#2{%
  \clap{$\mathsurround=0pt#1{#2}$}}

\let\oldstackrel\stackrel
\renewcommand{\stackrel}[2]{\oldstackrel{\mathclap{#1}}{#2}}

\newtheorem{theorem}{Theorem}
\newtheorem{definition}{Definition}
\newtheorem{lemma}{Lemma}
\newtheorem{proposition}{Proposition}

\newtheorem{remark}{Remark}

\newcommand{\zerov}{\mathbf{0}}
\newcommand{\Fq}{\mathbb{F}_q}
\newcommand{\Fqh}{\hat{\mathbb{F}}_q}

\newcommand{\Gb}{\mathsf{G}}

\newcommand{\Mc}{\mathcal{M}}

\newcommand{\Rrmac}{\Rr_\mathrm{MAC}}
\newcommand{\Rrc}{\Rr_\mathrm{CF}}

\newcommand{\Rrbc}{\Rr_\mathrm{BC}}
\newcommand{\Rrbci}{\Rr_{\mathrm{BC},1}}
\newcommand{\Rrbcii}{\Rr_{\mathrm{BC},2}}
\newcommand{\Rrbcj}{\Rr_{\mathrm{BC},j}}
\newcommand{\tRr}{\tilde{\Rr}}

\newcommand{\vf}{\varphi}
\newcommand{\sbq}{\subseteq} 

\newcommand{\Et}{{\tilde{E}}}
\newcommand{\Ft}{{\tilde{F}}}
\newcommand{\cl}{\mathrm{cl}}
\newcommand{\ab}{\overline{\a}}

\IEEEoverridecommandlockouts
\begin{document}

\title{On the Optimal Achievable Rates for Linear Computation with Random Homologous Codes}

\author{\IEEEauthorblockN{Pinar Sen,
						  Sung Hoon Lim,
						  and Young-Han Kim}%
\thanks{This work was supported in part by the Electronics and Telecommunications Research Institute through Grant 17ZF1100 from the Korean Ministry of Science, ICT, and Future Planning and in part by the National Research Foundation (NRF) of Korea  funded by the Ministry of Education, Science and Technology under Grant NRF-2017R1C1B1004192.}%
\thanks{This work was presented in part at the 2018 IEEE International Symposium of Information Theory, Vail, Colorado, June 17--22, 2018.}%
\thanks{Pinar Sen and Young-Han Kim are with the Department of Electrical and Computer Engineering, University of California, San Diego, La Jolla, CA 92093 USA (email: psen@ucsd.edu, yhk@ucsd.edu)}%
\thanks{Sung Hoon Lim is with the Korea Institute of Ocean Science and Technology, Busan 49111, Korea (email: shlim@kiost.ac.kr)}
}

\maketitle

% \fi
\IEEEpeerreviewmaketitle

\begin{abstract} 
The problem of computing a linear combination of sources over a multiple access channel is studied. Inner and outer bounds on the optimal tradeoff between the communication rates are established when encoding is restricted to random ensembles of \emph{homologous codes}, namely, structured nested coset codes from the same generator matrix and individual shaping functions,
but when decoding is optimized with respect to the realization of the encoders.
For the special case in which the desired linear combination is ``matched'' to the structure
of the multiple access channel in a natural sense, these inner and outer bounds coincide.
This result indicates that most, if not all, coding schemes for computation in the literature that rely on random construction of nested coset codes cannot be improved by using more powerful decoders, such as
the maximum likelihood decoder. The proof techniques are adapted to characterize the rate region for broadcast channels achieved by Marton's (random) coding scheme under
maximum likelihood decoding.
\end{abstract}

\section{Introduction}
\label{sec:intro}
\begin{figure}[t]
\center
\includegraphics[scale=0.85]{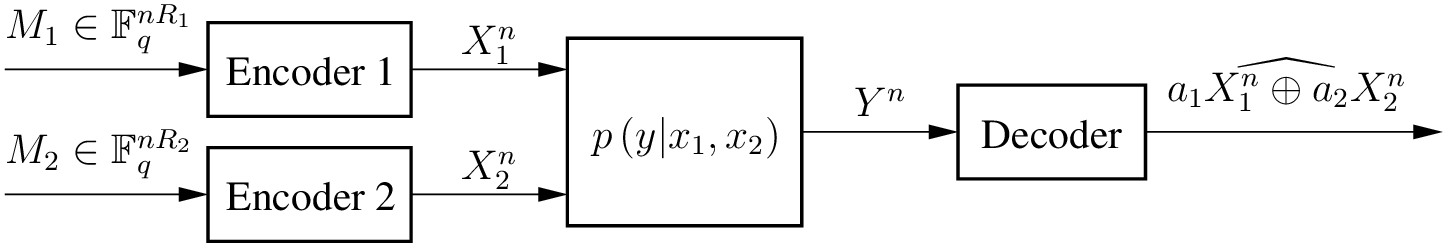}
\caption{Linear computation over two-sender multiple access channel}
\label{fig:comp}
\end{figure}

Consider a multiple access channel (MAC) with two senders and one receiver, in which the receiver wishes to 
reliably estimate a linear function of the transmitted sources
from the senders (see Figure~\ref{fig:comp}). One trivial approach to this \emph{computation} problem involves two steps: first recover the individual sources and then compute the function from the recovered sources. 
When the problem is isolated to the first communication step of this plug-in approach, using the conventional random independently and identically distributed (i.i.d.) code ensembles achieves the optimal rates of communicating independent sources~\cite{Ahlswede1971,Liao1972}. For the problem as a whole, however,
the use of random i.i.d.\@ code ensembles is strictly suboptimal even for a trivial MAC.
As shown by K\"orner and Marton~\cite{Korner--Marton1979}
for the problem of encoding a modulo-two sum of distributed dependent binary sources, 
using the \emph{same} random ensemble of linear codes at multiple encoders can achieve strictly better rates than using independently generated ensembles of codes.
Building on this observation, Nazer and Gastpar~\cite{Nazer--Gastpar2007a} developed a channel coding scheme that uses the same random ensemble of lattice codes at multiple encoders and showed that
this \emph{structured} coding scheme outperforms conventional random coding schemes 
for computing a linear combination of the sources over a linear MAC, even for independent sources.
This influential work led to the development of the \emph{compute--forward} strategy for relay networks~\cite{Wilson--Narayanan--Pfister--Sprintson2010,Nam--Chung--Lee2010,Nazer--Gastpar2011}. 
Over the past decade, the compute--forward strategy based on lattice codes and its extensions have shown to provide higher achievable rates for several communication problems over relay networks~\cite{Wilson--Narayanan--Pfister--Sprintson2010,Nam--Chung--Lee2010,Nazer--Gastpar2011,Niesen--Whiting2012,Song--Devroye2013,Hong--Caire2013,Ren--Goseling--Weber--Gastpar2014}.

More recently, \emph{nested coset codes}~\cite{Miyake2010, Padakandla--Pradhan2013c}
were proposed as more flexible alternatives for achieving the desired linear structure at multiple encoders.
In particular, Padakandla and Pradhan~\cite{Padakandla--Pradhan2013c} developed a fascinating coding scheme
for the computation problem over an \emph{arbitrary} MAC. 
In this coding scheme, a coset code with a rate higher than the target (message) rate is first generated randomly.
Next, in the \emph{shaping} step, a codeword of a desired property (such as type or joint type) is selected from a subset of codewords (a coset of a subcode). Although reminiscent of the
multicoding scheme of Gelfand and Pinsker~\cite{Gelfand--Pinsker1980a} for channels with state, and Marton's coding scheme~\cite{Marton1979} for broadcast channels, this construction is more fundamental in some sense, since the scheme is directly applicable even for classical point-to-point communication channels.
A similar shaping technique was also developed for lattice codes in~\cite{Tal--Erez2008}.
For multiple encoders, the desired common structure is obtained by using coset codes with the same generator matrix. Recent efforts exploited the benefit of such constructions for a broader class of channel models, such as interference channels~\cite{Padakandla--Pradhan2012,Padakandla--Pradhan2016}, multiple access channels~\cite{Sen--Kim2017,Sen--Kim2018s}, and multiple access channels with state~\cite{Padakandla--Pradhan2013}.

To develop a unified framework for the compute--forward strategy, Lim, Feng, Pastore, Nazer, and Gastpar~\cite{Lim--Gastpar2016,Lim--Gastpar2017} generalized the nested coset codes of the same generator matrix to asymmetric rate pairs. We referred to this generalized version, together with the shaping step, as \emph{homologous} codes~\cite{Sen--Kim2017, Sen--Kim2018s, Sen--Lim--Kim2018c}. This terminology is motivated from its biological definition, i.e., the structures modified from the same ancestry (underlying linear code) to adapt to different purposes (desired shape). Lim et al.~\cite{Lim--Gastpar2016,Lim--Gastpar2017}
further analyzed
\emph{simultaneous decoding} of random ensembles of homologous codes
and showed that it can achieve rates higher than existing approaches
to computation problems.
For instance, when adapted to the Gaussian MAC, the resulting achievable rates improve upon those of lattice codes~\cite{Nazer--Gastpar2011}.
With mathematical rate expressions in single-letter mutual information terms and with physical rate performances
better than those of lattice codes,
homologous codes have a potential to bringing a deeper understanding of the fundamental limits of the computation problem. 

Several open questions remain, however. What is the optimal tradeoff between achievable rates for reliable computation? Which scheme achieves this computation capacity region? The answers require a joint optimization of encoder and decoder designs, which seems to be
intractable as in many other network information theory problems.

In this paper, we instead concentrate on the performance of the optimal maximum likelihood decoder when the encoder is restricted to a given random ensemble of homologous codes. We characterize the optimal rate region when the desired linear combination and the channel structure are ``matched'' (see Definition~\ref{def:natural} in Section~\ref{sec:main_result}), which is the case in which the benefit of computation can be realized to the fullest extent as indicated by~\cite{Karamchandani--Niesen--Diggavi2013}. This result, inter alia,  implies that the suboptimal joint typicality decoding rule proposed in~\cite{Lim--Gastpar2016,Lim--Gastpar2017} achieves this optimal rate region. Thus, the performance of random ensembles of homologous codes cannot be improved by the maximum likelihood decoder. 

The main contribution lies in the outer bound on the optimal rate region (Theorem~\ref{thm:outer_ncc}), which characterizes the necessary condition that a rate pair must satisfy if the average probability of decoding error vanishes asymptotically.
The proof of this bound relies on two key observations. First, the distribution of a given random ensemble of homologous codes converges asymptotically to the product of the desired input distribution. Second, given the channel output, a relatively short list of messages can be constructed that includes the actually transmitted message with high probability. 
The second observation, which is adapted
from the analysis in~\cite{Bandemer--El-Gamal--Kim2012a} for the optimal rate region of interference networks with random i.i.d. code ensembles, seems to be a recurring path to establishing the optimal performance of random code ensembles.

As hinted earlier, the construction of random ensemble of homologous codes has many similarities to Marton's coding scheme~\cite{Marton1979}, one of the fundamental coding schemes in network information theory. As a result, adapting the proof techniques that we developed for homologous codes, we can establish an outer bound on the optimal rate region for broadcast channels with Marton's coding scheme (Proposition~\ref{prop:outer_marton}). The resulting outer bound coincides with the inner bound that is achieved by \emph{simultaneous nonunique decoding}, thus characterizing the optimal rate region of a two-receiver general broadcast channel achieved by a given random code ensemble.

The rest of the paper is organized as follows. Section~\ref{sec:prob_def} formally defines the computation problem. Section~\ref{sec:main_result} presents the main result of the paper---the optimal rate region achievable by a random ensemble of homologous codes. The inner and the outer bounds on this region are presented in Sections~\ref{sec:achiev} and \ref{sec:converse}, respectively. Section~\ref{sec:marton} discusses the optimal rate region for a broadcast channel achievable by Marton's coding scheme. 

We adapt the notation in~\cite{Cover--Thomas2006, El-Gamal--Kim2011}.
The set of integers $\{ 1,2,\ldots, n \}$ is denoted by $[n]$.
For a length-$n$ sequence (row vector) $x^n=(x_1,x_2,\ldots,x_n) \in \Xc^n$, we define its type
as $\pi(x | x^n) = {|\{ i \suchthat x_i = x \}|}/{n}$ for $x \in \Xc$. Upper case letters $X,Y,\ldots$ denote random variables. 
For $\e \in (0,1)$, we define the $\e$-typical set of $n$-sequences (or the typical set in short) as $\aep(X) = \{ x^n \suchthat | p(x) - \pi(x| x^n) | \le \e p(x), \, x \in \Xc \}$. The indicator function $\mathbbm{1}_{\Sc}: \Xc \to \{ 0, 1 \}$ for $\Sc \sbq \Xc$ is
defined as $\mathbbm{1}_{\Sc}(x) = 1$ if $x \in \Sc$ and $0$ otherwise. A length-$n$ row vector of all zeros is denoted by $\mathbf{0}_n$, where the subscript is omitted when it is clear from the context. We denote by $\Fq$ a finite field of size $q$, $\Fq^*$ is the set of nonzero elements in $\Fq$, and $\Fq^{d}$ is the $d$-dimensional vector space over $\Fq$. The limit of a collection of sets $\{\Ac(\e)\}$ indexed by $\epsilon>0$ is defined as
\begin{equation}
\label{eq:limit_reg}
\lim_{\e \to 0} \Ac(\e) := 
\bigcup_{\e > 0} \; \bigcap_{0 < \g < \e} \Ac(\g) \overset{(a)}{=}
\bigcap_{\e > 0} \; \bigcup_{0 < \g < \e} \Ac(\g),
\end{equation}
which exists if $(a)$ holds. The closure $\cl(\Ac)$ of a set $\Ac \subseteq \Real^d$ denotes the smallest closed superset of $\Ac$. We use $\e_n \ge 0$ to denote a generic sequence of $n$ that tends to zero as $n \to \infty$, and use $\d_i(\e) \ge 0$, $i \in \Zz^+$, to denote a continuous function of $\e$ that tends to zero as $\e \to 0$. Throughout the paper, information measures are in logarithm base~$q$.

\section{Formal Statement of the Problem}
\label{sec:prob_def}
Consider the two-sender finite-field input memoryless multiple access channel (MAC)
\[ 
(\Xc_1\times\Xc_2, p(y|x_1, x_2), \Yc)
\]
in Figure~\ref{fig:comp}, which consists of two sender alphabets $\Xc_1 = \Xc_2 = \Fq$, a receiver alphabet $\Yc$, and a collection of conditional probability distributions $p_{Y|X_1,X_2}(y|x_1,x_2)$.
Each sender $j=1,2$ encodes a message $M_j \in \Fq^{nR_j}$ into a codeword $X_j^n = x_j^n(M_j) \in \Fq^n$ and transmits $X_j^n$ over the channel. Here and henceforth, we assume without loss of generality that $nR_1$ and $nR_2$ are integers. The goal of communication is to convey a linear combination of the codewords. Hence, the receiver finds an estimate  $\Wh^n_{\av} = \wh_{\av}^n(Y^n) \in \Fq^n$ of
\[
W_{\av}^n := a_1 X_1^n \oplus a_2 X_2^n
\] 
for a desired (nonzero) vector $\av = [a_1 \; a_2]$ over $\Fq$.
Formally, an $(n,nR_1,nR_2)$ \emph{computation code} for the multiple access channel consists of two encoders that map $x_j^n(m_j)$, $j=1,2$, and a decoder that maps $\wh^n_{\av}(y^n)$. The collection of codewords $\cc_n := \{(x_1^n(m_1), x_2^n(m_2)): (m_1,m_2) \in \Fq^{(nR_1) \times (nR_2)}\}$ is referred to as the \emph{codebook} associated with the $(n,nR_1,nR_2)$ code.

\begin{remark}
For simplicity of presentation, we consider the case $\Xc_1 = \Xc_2 = \Fq$, but our arguments can be extended to arbitrary $\Xc_1$ and $\Xc_2$ through the channel transformation technique by Gallager~\cite[Sec. 6.2]{Gallager1968}. More specifically, given a pair of symbol-by-symbol mappings $\vf_j: \Fq \to \Xc_j$, $j=1,2$, consider the \emph{virtual channel} with finite field inputs, $p(y | v_1,v_2) = p_{Y | X_1,X_2}(y | \vf_1(v_1),\vf_2(v_2))$, for which a computation code is to be defined. The goal of the communication is to convey $W_{\av} := a_1 V_1^n \oplus a_2 V_2^n$, where $V_j^n = v_j^n(M_j) \in \Fq^n$ is the virtual codeword mapped to message $M_j$ at sender $j=1,2$. Our results can be readily applied to this computation problem defined on the virtual channel.  
\end{remark}

The performance of a given computation code with codebook $\cc_n$ is measured by the average probability of error
\[
P_e^{(n)} (\cc_n) = \P( \Wh_{\av}^n \neq W_{\av}^n | \cc_n),
\]
when $M_1$ and $M_2$ are independent and uniformly distributed. 
A rate pair $(R_1,R_2)$ is said to be \emph{achievable} if there exists a sequence of $(n,nR_1,nR_2)$ computation codes such that 
\[
\lim_{n \to \infty} P_e^{(n)} (\cc_n) = 0
\]
and
\begin{equation}
\label{eq:cond_ent}
\lim_{n \to \infty} H(M_j | x_j^n(M_j), \cc_n) = 0, \quad j \in \{1,2\}  \textrm{ with }  a_j \neq 0.
\end{equation}
Note that without the condition in (\ref{eq:cond_ent}), the problem is trivial and an arbitrarily large rate pair is achievable. 

We now define the random ensemble of computation codes referred to as homologous codes. Let $p = p(x_1) p(x_2)$ be a given input pmf on $\Fq \times \Fq$, and let $\e > 0$. Suppose that the codewords $x_1^n(m_1)$, $m_1 \in \Fq^{nR_1}$, and $x_2^n(m_2)$, $m_2 \in \Fq^{nR_2}$ that constitute the codebook are generated according to the following steps:
\begin{enumerate}
\item Let $\Rh_j = D(p_{X_j} \| \U(\Fq))+\e$, $j=1,2,$ where $D(\cdot\|\cdot)$ is the Kullback--Leibler divergence. %Assume without loss of generality that $n \Rh_1$ and $n \Rh_2$ are integers.
\item Randomly generate a $\kappa \times n$ generator matrix $G$, and two dither vectors $D_1^n$ and $D_2^n$ such that the elements of $G, D_1^n$, and $D_2^n$ are i.i.d. $\U(\Fq)$ random variables, where $\kappa = \max \{ nR_1  +  n \Rh_1, nR_2 + n \Rh_2 \}$.
\item Given the realizations $\Gb, d_1^n$, and $d_2^n$ of the generator matrix and dithers, let 
\[
u_j^n(m_j,l_j) = [m_j \;\: l_j \;\: \zerov] \: \Gb + d_j^n, \quad m_j \in \Fq^{nR_j}, \; l_j \in \Fq^{n\Rh_j}, \; j=1,2.
\]
At sender $j=1,2$, assign a codeword $x_j^n(m_j) = u_j^n(m_j, L_j(m_j))$ to each message $m_j \in \Fq^{nR_j}$ where $L_j(m_j)$ is a random variable that is drawn uniformly at random among all $l_j$ vectors satisfying $u_j^n(m_j,l_j) \in \aep(X_j)$ if there exists one, or among $\Fq^{n\Rh_j}$ otherwise. 
\end{enumerate}
With a slight abuse of terminology, we refer to the random tuple $\Cc_n := (G, D_1^n, D_2^n, (L_1(m_1): m_1 \in \Fq^{nR_1}), (L_2(m_2): m_2 \in \Fq^{nR_2}))$ as the \emph{random homologous codebook}. Each realization of the random homologous codebook $\Cc_n$ results in one instance $\{(x_1^n(m_1), x_2^n(m_2)): (m_1,m_2) \in \Fq^{nR_1} \times \Fq^{nR_2} \}$ of such generated codebooks, which constitutes an $(n,nR_1,nR_2)$ computation code along with the optimal decoder. The random code ensemble generated in this manner is referred to as an $(n, nR_1,nR_2; p, \e)$ \emph{random homologous code ensemble}, where $p$ is the given input pmf and $\e>0$ is the parameter used in steps $1$ and $3$ in codebook generation. A rate pair $(R_1,R_2)$ is said to be \emph{achievable by the $(p,\e)$-distributed random homologous code ensemble} if there exits a sequence of $(n, nR_1,nR_2; p, \e)$ random homologous code ensembles such that
\[ 
\lim_{n \to \infty} \E_{\Cc_n} [ P_e^{(n)} (\Cc_n)] = 0
\]
and
\begin{equation}
\label{eq:con_ent_avg}
\lim_{n \to \infty} H(M_j | X_j^n(M_j), \Cc_n) = 0, \quad j \in \{1,2\}  \textrm{ with }  a_j \neq 0.
\end{equation}
Here the expectation is with respect to the random homologous codebook $\Cc_n$, i.e., $(G, D_1^n, D_2^n, (L_1(m_1): m_1 \in \Fq^{nR_1}), (L_2(m_2): m_2 \in \Fq^{nR_2}))$. Given $(p,\e)$, let $\Rr^*(p,\e)$ be the set of all rate pairs achievable by the $(p,\e)$-distributed random homologous code ensemble. Given the input pmf $p$, the optimal rate region $\Rr^*(p)$, when it exists, is defined as
\[
\Rr^*(p) := \cl \left[\lim_{\e \to 0} \Rr^*(p,\e) \right].
\]

\section{Main Result}
\label{sec:main_result}
In this section, we present a single-letter characterization of the optimal rate region when the target linear combination is in the following class.
\begin{definition}
\label{def:natural}
A linear combination $W_{\av} = a_1 X_1 \oplus a_2 X_2$ for some $\av = [a_1 \; a_2]  \in \Fq^2 \setminus{ \{ \zerov\}}$ is said to be \emph{natural} if 
\begin{align}
\label{eq:comp_favor}
H(W_{\av}|Y) = \min_{\bv \neq \zerov} H(W_{\bv} |Y),
\end{align}
where $\bv = [b_1 \; b_2]$ and $W_{\bv} = b_1 X_1 \oplus b_2 X_2$ are over $\Fq$. 
\end{definition}

In words, a natural combination $W_{\av}$ is the easiest to recover at the receiver and thus, in some sense, is the best linear combination that is matched to the channel structure.

We are now ready to present the optimal rate region for computing natural linear combinations.

\begin{theorem}
\label{thm:optimal_comp}
Given an input pmf $p = p(x_1)p(x_2)$, the optimal rate region $\Rr^*(p)$ for computing a natural combination $W_{\av}$ is the set of rate pairs $(R_1,R_2)$ such that
\begin{subequations}
\label{eq:Rr**}
\begin{align}
R_j &\le I(X_j; Y|X_{j^c}),\\
R_j &\le I(X_1, X_2; Y)-\min\{R_{j^c}, I(X_{j^c}; W_{\av}, Y)\}
\end{align}
\end{subequations}%
for every $j \in \{1,2\}$ with $a_j \neq 0$, where $j^c = \{1,2 \} \setminus \{j \}$.
\end{theorem}

The rate region in (\ref{eq:Rr**}) in Theorem \ref{thm:optimal_comp}, which we will denote as $\Rr^{**}(p)$, can be equivalently characterized in terms of well-known rate regions for compute--forward and message communication.
Let $\Rrc(p)$ be the set of rate pairs $(R_1,R_2)$ such that 
\begin{equation}
\label{eq:Rcf}
R_j \le H(X_j) - H(W_{\av} | Y), \quad \forall j \in \{1,2\} \text{ with } a_j \neq 0.
\end{equation}
Let $\Rrmac(p)$ be the set of rate pairs $(R_1,R_2)$ such that
\begin{align*}
R_1 &\le I(X_1;Y|X_2), \\
R_2 &\le I(X_2;Y|X_1), \\
R_1 + R_2 &\le I(X_1,X_2;Y).
\end{align*} 

\begin{proposition}
\label{prop:opt_reg}
For any input pmf $p=p(x_1)p(x_2)$ and any linear combination $W_{\av}$, 
\[
\Rr^{**}(p) = \Rrc(p) \cup \Rrmac(p).
\]
\end{proposition}

The proof of Proposition~\ref{prop:opt_reg} is relegated to Appendix~\ref{app:prop1}.

We prove Theorem \ref{thm:optimal_comp} in three steps: 1) we first present a general (not necessarily for natural combinations) inner bound on the optimal rate region in Section~\ref{sec:achiev}, where we follow the results in~\cite{Lim--Gastpar2016, Lim--Gastpar2017} that studied the rate region achievable by random homologous code ensembles using a suboptimal joint typicality decoding rule, 2) we then show by Lemma \ref{lem:equiv_reg} in Section~\ref{sec:achiev} that this inner bound is equivalent to $\Rr^{**}(p)$ in Proposition \ref{prop:opt_reg} if $W_{\av}$ is a natural combination, and 3) we present a general (not necessarily for natural combinations) outer bound on the optimal rate region in Section~\ref{sec:converse} by showing that if a rate pair $(R_1,R_2)$ is achievable by the $(p,\e)$-distributed random homologous code ensemble for arbitrarily small $\e$, then $(R_1,R_2)$ must lie in $\Rr^{**}(p)$ in Theorem \ref{thm:optimal_comp}.

\section{An Inner Bound}
\label{sec:achiev}
The computation performance of random homologous code ensembles was studied using a suboptimal \emph{joint typicality} decoder in~\cite{Lim--Gastpar2016, Lim--Gastpar2017}. For completeness, we first describe the joint typicality decoding rule and then characterize the rate region achievable by the $(p,\e)$-distributed random homologous code ensemble \emph{under this joint typicality decoding rule}. We then concentrate on an arbitrarily small $\e$ to provide an inner bound on the optimal rate region $\Rr^*(p)$. We will omit the steps that were already established in~\cite{Lim--Gastpar2016, Lim--Gastpar2017} and instead provide detailed references.

Upon receiving $y^n$, the $\e'$-joint typicality decoder, $\e'>0$, looks for a unique vector $s \in \Fq^{\kappa}$ such that
\[
s = a_1[m_1 \; l_1\; \mathbf{0}] \oplus a_2 [m_2 \; l_2\; \mathbf{0}],
\]
for some $(m_1,l_1,m_2,l_2) \in \Fq^{nR_1} \times \Fq^{n\Rh_1} 
\times \Fq^{nR_2} \times \Fq^{n\Rh_2}$ that satisfies
\[
(u_1^n(m_1,l_1), u_2^n(m_2,l_2),y^n) \in \aepvar(X_1,X_2,Y).
\]
If the decoder finds such $s$, then it declares $\wh_{\av}^n = s \Gb \oplus a_1 d_1^n \oplus a_2 d_2^n$ as an estimate; otherwise, it declares an error.

To describe the performance of the joint typicality decoder, we define $\Rrc(p,\d)$ for a given input pmf $p$ and $\d \ge 0$ as the set of rate pairs $(R_1,R_2)$ such that 
\[
R_j \le H(X_j) - H(W_{\av} | Y) - \d, \quad \forall j \in \{1,2\} \text{ with } a_j \neq 0.
\]
Similarly, we define $\Rr_1(p,\d)$ as the set of rate pairs $(R_1,R_2)$ such that
\begin{subequations}
\label{eq:Rr1}
\begin{align}
R_1 &\le I(X_1;Y|X_2) - \d, \\
R_2 &\le I(X_2;Y|X_1) - \d, \\
R_1 + R_2 &\le I(X_1,X_2;Y) - \d, \\
R_1 &\le I(X_1,X_2;Y) - H(X_2) + \min_{b_1,b_2 \in \Fq^*} H(W_{\bv}|Y) - \d,
\end{align}
\end{subequations}%
and $\Rr_2(p,\d)$ as the set of rate pairs $(R_1,R_2)$ such that
\begin{subequations}
\label{eq:Rr2}
\begin{align}
R_1 &\le I(X_1;Y|X_2) - \d, \\
R_2 &\le I(X_2;Y|X_1) - \d, \\
R_1 + R_2 &\le I(X_1,X_2;Y) - \d, \\
R_2 &\le I(X_1,X_2;Y) - H(X_1) + \min_{b_1,b_2 \in \Fq^*} H(W_{\bv}|Y) - \d,
\end{align}
\end{subequations}%
where $\bv = [b_1 \; b_2]$ and $W_{\bv} = b_1 X_1 \oplus b_2 X_2$ are over $\Fq$. Note that the region $\Rrc(p) = \Rrc(p,\d=0)$, as defined in (\ref{eq:Rcf}) in Section~\ref{sec:main_result}. Similarly, let $\Rr_{j}(p)$ denote the region $\Rr_{j}(p,\d=0)$ for $j=1,2$ in (\ref{eq:Rr1}) and (\ref{eq:Rr2}).

We are now ready to state the rate region achievable by the random homologous code ensembles that combines the inner bounds in~{\cite[Theorem 1]{Lim--Gastpar2016} and~\cite[Corollary 1]{Lim--Gastpar2017}.

\begin{theorem}
\label{thm:achiev}
Let $p=p(x_1)p(x_2)$ be an input pmf and $\d>0$. Then, there exists $\e' < \d$ such that for every $\e < \e'$ sufficiently small, a rate pair 
\begin{equation}
\label{eq:comp_inner}
(R_1,R_2) \in \Rr_{CF}(p,\d) \cup \Rr_1(p,\d) \cup \Rr_2(p,\d)
\end{equation}
is achievable by the $(p,\e)$-distributed random homologous code ensemble along with the $\e'$-joint typicality decoder for computing an \emph{arbitrary} linear combination $W_{\av}$.
In particular,
\begin{equation}
\label{eq:comp_inner_limit}
[\Rr_{CF}(p) \cup \Rr_1(p) \cup \Rr_2(p)] \sbq \Rr^{*}(p).
\end{equation}
\end{theorem}

\begin{IEEEproof}
By~\cite[Theorem 1]{Lim--Gastpar2016}, for sufficiently small $\e < \e' < \d$, the average probability of error for the $(p,\e)$-distributed random homologous code ensemble paired with the $\e'$-joint typicality decoder tends to zero as $n \to \infty$ if   
\begin{equation}
\label{eq:achiev_cond1}
(R_1,R_2) \in \Rr_{CF}(p,\d).
\end{equation}
Similarly, by~\cite[Corollary 1]{Lim--Gastpar2017}, the average probability of error tends to zero as $n \to \infty$ if 
\begin{equation}
\label{eq:achiev_cond2}
(R_1,R_2) \in \Rr_{1}(p,\d) \cup \Rr_{2}(p,\d).
\end{equation}
Combining (\ref{eq:achiev_cond1}) and (\ref{eq:achiev_cond2}) establishes (\ref{eq:comp_inner}).

We still need to show that the condition in (\ref{eq:con_ent_avg}) holds. Suppose that $a_j \neq 0$. Let $G_j$ denote the submatrix that consists of the first $(nR_j + n\Rh_j )$ rows of $G$ and $S_j$ be the indicator variable such that $S_j = 1$ if $G_j$ is full rank. Then,
\begin{align*}
H(M_j|X_j^n(M_j),\Cc_n) &= H(M_j | X_j^n(M_j), 
G, D_1^n, D_2^n, (L_1(m_1): m_1 \in \Fq^{nR_1}),(L_2(m_2): m_2 \in \Fq^{nR_2})) \\
& \le H(M_j | X_j^n(M_j), 
G, D_j^n, (L_j(m_j): m_j \in \Fq^{nR_j})) \\
&= H(M_j | X_j^n(M_j), 
G, S_j, D_j^n, (L_j(m_j): m_j \in \Fq^{nR_j})) \\
&= H(M_j | X_j^n(M_j), 
G, S_j = 1, D_j^n, (L_j(m_j): m_j \in \Fq^{nR_j}))
\P(S_j = 1)  \\
& \quad \quad \quad \quad + H(M_j | X_j^n(M_j), 
G, S_j = 0, D_j^n, (L_j(m_j): m_j \in \Fq^{nR_j}))
\P(S_j = 0) 
\\
&=  H(M_j | X_j^n(M_j), 
G, S_j = 0, D_j^n, (L_j(m_j): m_j \in \Fq^{nR_j}))
\P(S_j = 0) \\
&\le  nR_j \P(S_j = 0).
\end{align*}
Now, by Lemma~\ref{lem:full_rank_G} in Appendix~\ref{app:full_rank} (with $R \leftarrow R_j + \Rh_j$), the term $n \P(S_j = 0)$ tends to zero as $n \to \infty$ if $R_j < H(X_j) - \e$. Since this condition is satisfied if (\ref{eq:comp_inner}) holds, the proof of (\ref{eq:comp_inner}) follows. 

The proof of (\ref{eq:comp_inner_limit}) follows by taking the closure of the union of (\ref{eq:comp_inner}) over all $\d>0$, which completes the proof of Theorem~\ref{thm:optimal_comp}.
\end{IEEEproof}

The inner bound (\ref{eq:comp_inner_limit}) in Theorem~\ref{thm:optimal_comp} is valid for computing an arbitrary linear combination, which may not be equal to the rate region $\Rr^{**}(p)$ in Theorem \ref{thm:optimal_comp} in general. For computing a \emph{natural} linear combination, however, the following lemma shows that the equivalent rate region in Proposition \ref{prop:opt_reg} is achievable.

\begin{lemma}
\label{lem:equiv_reg}
If the desired linear combination $W_{\av}=a_1 X_1 \oplus a_2 X_2$ for $(a_1,a_2) \neq (0,0)$ is natural, then 
\[
[\Rrc(p) \cup \Rr_{1}(p) \cup \Rr_2(p)] = [\Rrc(p) \cup \Rrmac(p)].
\]
\end{lemma}

The proof of Lemma~\ref{lem:equiv_reg} is relegated to Appendix~\ref{app:lemma1}.

\section{An Outer Bound}
\label{sec:converse}
We first present an outer bound on the rate region $\Rr^*(p,\e)$ for a fixed input pmf $p$ and $\e>0$. We then discuss the limit of this outer bound as $\e \to 0$ to establish an outer bound on the rate region $\Rr^*(p)$. Given an input pmf $p$ and $\d >0$, we define the rate region $\Rr^{**}(p,\d)$ as the set of rate pairs $(R_1,R_2)$ such that
\begin{subequations}
\begin{align}
\label{eq:type1}
R_j &\le I(X_j; Y|X_{j^c}) + \d,\\
\label{eq:type2}
R_j &\le I(X_1, X_2; Y)-\min\{R_{j^c}, I(X_{j^c}; W_{\av}, Y)\} + \d,
\end{align}
\end{subequations}%
for every $j \in \{1,2\}$ with $a_j \neq 0$, where $j^c = \{1,2 \} \setminus \{j \}$. Note that $\Rr^{**}(p, \d=0)$ is equal to $\Rr^{**}(p)$ as defined in (\ref{eq:Rr**}).

We are now ready to state the outer bound on the optimal rate region for computing an \emph{arbitrary} linear combination, which is also an outer bound on $\Rr^*(p)$ in Theorem \ref{thm:optimal_comp} for computing a \emph{natural} combination.

\begin{theorem}
\label{thm:outer_ncc}
Let $p=p(x_1)p(x_2)$ be an input pmf and $\e>0$. If a rate pair $(R_1, R_2)$ is achievable by the $(p,\e)$-distributed random homologous code ensemble for computing an arbitrary linear combination $W_{\av}$, then there exists a continuous $\d'(\e)$ that tends to zero monotonically as $\e \to 0$ such that
\begin{equation}
\label{eq:outer_comp}
(R_1,R_2) \in \Rr^{**}(p,\d'(\e)).
\end{equation}
In particular,
\begin{equation}
\label{eq:outer_comp_limit}
\Rr^*(p) \sbq \Rr^{**}(p).
\end{equation}
\end{theorem}

\begin{IEEEproof}
We first start with an averaged version of Fano's inequality for a random homologous code ensemble $\Cc_n$ (recall the notation in Section~\ref{sec:prob_def}), the proof of which is relegated to Appendix~\ref{app:fano_homolog}.
\begin{lemma}
\label{lem:Fano_homolog}
If 
\[
\lim_{n \to \infty} \E_{\Cc_n} [ P_e^{(n)}(\Cc_n)] = 0
\]
and
\[
\lim_{n \to \infty} H(M_j | X_j^n(M_j), \Cc_n) = 0
\]
for $j \in \{1,2\}$ with $a_j \neq 0$, then for each $j \in \{1, 2\}$ with $a_j \neq 0$
\[
H(M_j|Y^n,M_{j^c},\Cc_n) \le n \e_n
\]
for some $\e_n \to 0$ as $n \to \infty$.
\end{lemma} 

We next define the indicator random variable 
\begin{equation}
\label{eq:indicator_en}
E_n = \mathbbm{1}_{\{ (X_1^n(M_1),X_2^n(M_2)) \in \aepvar(X_1,X_2) \}}
\end{equation}%
for $\e'>0$. Since $\Rh_i = D(p_{X_i} \| \U(\Fq)) + \e $, $i=1,2$, by the Markov lemma~\cite[Lemma 12]{Lim--Gastpar2016} for homologous codes, $\P(E_n = 0)$ tends to zero as $n \to \infty$ if $\e'$ is sufficiently large compared to $\e$. Let $\e' = \d_1(\e)$, which still tends to zero as $\e \to 0$. Suppose that $a_j \neq 0$. 
Then, for $n$ sufficiently large, 
\begin{align} \notag
nR_j &= H(M_j|M_{j^c}, \Cc_n) 
\\ \notag
& \overset{(a)}{\le} I(M_j;Y^n| M_{j^c}, \Cc_n) + n \e_n 
\\ \notag
& \le I(M_j, E_n ;Y^n| M_{j^c}, \Cc_n) + n \e_n 
\\ \notag
& \overset{(b)}{\le} 1 + I(M_j ;Y^n| M_{j^c}, \Cc_n, E_n) + n \e_n 
\\ \notag 
& \le 1 + I(M_j ;Y^n| M_{j^c}, \Cc_n, E_n = 0) \P(E_n = 0) + I(M_j ;Y^n| M_{j^c}, \Cc_n, E = 1) \P(E_n = 1) + n \e_n 
\\ \notag
& \le 1 + nR_j \P(E_n = 0) + I(M_j ;Y^n| M_{j^c}, \Cc_n, E_n = 1) + n \e_n 
\\ \notag
& = 1 + nR_j \P(E_n = 0) + \sum_{i=1}^{n} I(M_j;Y_i | Y^{i-1}, M_{j^c}, \Cc_n, X_{j^c i}, E_n = 1) + n \e_n 
\\ \notag
& \le 1 + nR_j \P(E_n = 0) + \sum_{i=1}^{n} I(M_j, X_{ji}, Y^{i-1}, M_{j^c}, \Cc_n ;Y_i |  X_{j^c i}, E_n = 1) + n \e_n 
\\ 
&\overset{(c)}{=} 1 + nR_j \P(E_n = 0) + \sum_{i=1}^{n} I( X_{ji} ;Y_i |  X_{j^c i}, E_n = 1) + n \e_n,
\label{eq:outer0}
\end{align}%
where $(a)$ follows by Lemma~\ref{lem:Fano_homolog}, $(b)$ follows since $E_n$ is a binary random variable, and $(c)$ follows since $(M_1,M_2, Y^{i-1}, \Cc_n, E_n)\to(X_{1i}, X_{2i}) \to Y_i$ form a Markov chain for every $i \in [n]$. To further upper bound (\ref{eq:outer0}), we make a connection between the distribution of the random homologous codebook and the input pmf $p$ as follows.

\begin{lemma}
\label{lem:output_ncc_input}
Let $(X,Y) \sim p_{X,Y}(x,y)$ on $\Fq \times \Yc$ and $\e > 0$. Let $X^n(m)$ be the random codeword assigned to message $m \in \Fq^{nR}$ by an $(n,nR;p_X,\e)$ random homologous code ensemble. Further let $Y^n$ be a random sequence distributed according to $\prod_{i=1}^n p_{Y|X}(y_i|x_i)$. Then, for every $(x,y) \in \Fq \times \Yc$
\[
(1-\e) p_{X,Y}(x,y) \le \P(X_i = x, Y_i = y | X^n \in \aep(X)) \le (1+\e) p_{X,Y}(x,y),
\]
where $i=1,2,\ldots,n$. 
\end{lemma}

The proof of Lemma~\ref{lem:output_ncc_input} is relegated to Appendix~\ref{app:lemma_pmf}.

Back to the proof of Theorem~\ref{thm:outer_ncc}, we are now ready to establish (\ref{eq:type1}). Combining (\ref{eq:outer0}) with Lemma~\ref{lem:output_ncc_input} (with $p(x) \leftarrow p(x_1)p(x_2)$), we have
\begin{align} \notag
nR_j &\le 1 + nR_j \P(E_n = 0) + n ( I( X_j ;Y |  X_{j^c}) + \d_2(\e)) + n \e_n
\\
\label{eq:outer1}
& \overset{(d)}{\le} n ( I( X_j ;Y |  X_{j^c}) + \d_2(\e)) + 2 n \e_n,
\end{align}%
where $(d)$ follows since $\P(E_n = 0)$ tends to zero as $n \to \infty$.

For the proof of (\ref{eq:type2}), we start with
\begin{align} \notag
nR_j &= H(M_j|M_{j^c}, \Cc_n) 
\\ \notag
& \overset{(a)}{\le} I(M_j;Y^n| M_{j^c}, \Cc_n) + n \e_n 
\\
\label{eq:outer2_0}
& = I(M_1, M_2; Y^n|\Cc_n) - I(M_{j^c}; Y^n |\Cc_n) + n \e_n,
\end{align}%
where $(a)$ follows by Lemma~\ref{lem:Fano_homolog}.
Following arguments similar to (\ref{eq:outer1}), the first term in (\ref{eq:outer2_0}) can be bounded as
\begin{align} \notag
I(M_1, M_2; Y^n|\Cc_n) &\le 1 + n(R_1+R_2) \P(E_n = 0) + \sum_{i=1}^n I(M_1, M_2; Y_i | \Cc_n, Y^{i-1}, E_n=1)
\\ \notag
& \le n \e_n + \sum_{i=1}^n I(M_1, M_2, \Cc_n, Y^{i-1} ; Y_i | E_n = 1) 
\\ \notag
&= n \e_n + \sum_{i=1}^n I(M_1, M_2, \Cc_n, Y^{i-1}, X_{1i}, X_{2i} ; Y_i | E_n = 1) 
\\ \notag
&= n \e_n + \sum_{i=1}^n I(X_{1i}, X_{2i} ; Y_i | E_n = 1)
\\
\label{eq:outer2_1}
&\le n \e_n + n (I(X_1, X_2 ; Y) + \d_3(\e)).
\end{align}%

To bound the second term in (\ref{eq:outer2_0}), we need the following lemma, which is proved in Appendix~\ref{app:list_homolog}.
\begin{lemma}
\label{lem:outer2_list} 
For every $\e'' > \e'$ and for $n$ sufficiently large,
\[
I(M_{j^c}; Y^n |\Cc_n) \ge n [ \min \{ R_{j^c}, I(X_{j^c};W_{\av}, Y) \} - \d_4(\e'') ]- n \e_n.
\]
\end{lemma}

Combining (\ref{eq:outer2_0}), (\ref{eq:outer2_1}), and Lemma \ref{lem:outer2_list} with $\e'' = 2\d_1(\e)$, we have
\begin{align}
\label{eq:outer2}
nR_j \le n (I(X_1,X_2;Y) + \d_3(\e)) - n [\min \{ R_{j^c}, I(X_{j^c};W_{\av},Y) \} - \d_5(\e)] + 2 n \e_n
\end{align}%
for $n$ sufficiently large. Letting $n \to \infty$ in (\ref{eq:outer1}) and (\ref{eq:outer2}) establishes
\begin{align*}
R_j & \le I(X_j; Y | X_{j^c}) + \d_2(\e), 
\\
R_j & \le I(X_1,X_2;Y) - \min \{ R_{j^c}, I(X_{j^c};W_{\av},Y) \} + \d_6(\e).
\end{align*}
The proof of (\ref{eq:outer_comp}) follows by taking a continuous monotonic function $\d'(\e) \ge \max \{\d_2(\e), \d_6(\e) \}$ that tends to zero as $\e \to 0$. Letting $\e \to 0$ in (\ref{eq:outer_comp}) establishes (\ref{eq:outer_comp_limit}), which completes the proof of Theorem~\ref{thm:outer_ncc}.
\end{IEEEproof}

\section{Optimal Achievable Rates for Broadcast Channels with Marton Coding}
\label{sec:marton}
In this section, we apply the techniques developed in the previous sections to establish the optimal rate region for broadcast channels by Marton coding. Consider the two-receiver discrete memoryless broadcast channel (DM-BC) $(\Xc, p(y_1,y_2|x), \Yc_1 \times \Yc_2)$ in Fig.~\ref{fig:bc}, where the sender communicates independent messages $M_1$ and $M_2$ to respective receivers (see~\cite{Cover1972, Cover1998, Marton1979} for the formal definition of the communication problem over the broadcast channel).

\begin{figure}[h]
\center
\includegraphics[scale=0.85]{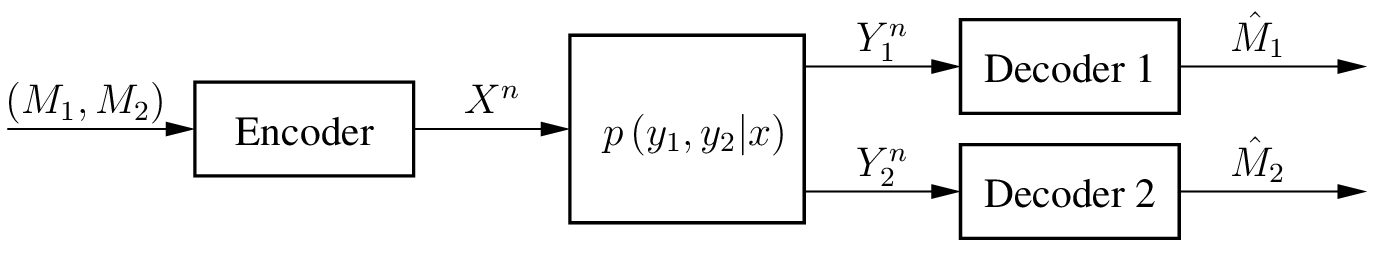}
\caption{Two-receiver broadcast channel}
\label{fig:bc}
\end{figure}

Let $p = p(u_1,u_2)$ be a given pmf on some finite set $\Uc_1 \times \Uc_2$, and $x(u_1,u_2)$ be a function from $\Uc_1 \times \Uc_2$ to $\Xc$, and let $\e > 0$ and $\a \in [0 \; 1]$. The random ensemble of \emph{Marton} codes \cite{Marton1979} is generated according to the following steps:
\begin{enumerate}
\item Let $\Rh_1 = \a(I(U_1;U_2) + 10 \e H(U_1,U_2))$ and $\Rh_2 = \ab(I(U_1;U_2) + 10 \e H(U_1,U_2))$, where $\ab := (1-\a)$. 

\item For each $m_1 \in [2^{nR_1}]$, generate \emph{auxiliary} codewords $u_1^n(m_1,l_1), l_1 \in [2^{n\Rh_1}]$, each drawn i.i.d. from $p(u_1)$. Similarly, for each $m_2 \in [2^{nR_2}]$, generate \emph{auxiliary} codewords $u_2^n(m_2,l_2), l_2 \in [2^{n\Rh_2}]$, each drawn i.i.d. from $p(u_2)$.

\item At the sender, for each message pair, $(m_1,m_2) \in [2^{nR_1}] \times [2^{nR_2}]$, find an index pair $(l_1,l_2) \in [2^{n\Rh_1}] \times [2^{n\Rh_2}]$ such that 
\[
(u_1^n(m_1,l_1), u_2^n(m_2,l_2)) \in \aep(U_1,U_2),
\]
and assign the codeword $x^n(m_1,m_2)$ as $x_i(m_1,m_2) = x(u_{1i}(m_1, l_1), u_{2i}(m_2, l_2)), i \in [n]$. If there are more than one such pair of $(l_1,l_2)$, choose one of them uniformly at random; otherwise, choose one uniformly at random from $[2^{n\Rh_1}] \times [2^{n\Rh_2}]$.
\end{enumerate}
We refer to the random tuple $\Cc_n := ((U_1^n(m_1,l_1): m_1 \in [2^{nR_1}], l_1 \in [2^{n\Rh_1}]), (U_2^n(m_2,l_2): m_2 \in [2^{nR_2}], l_2 \in [2^{n\Rh_2}]), ((L_1,L_2,x)(m_1,m_2): m_1 \in [2^{nR_1}], m_2 \in [2^{nR_2}]))$ as the \emph{Marton random codebook}.
Each realization of the Marton random codebook $\Cc_n$ results in one instance $\{x^n(m_1,m_2): (m_1,m_2) \in [2^{nR_1}] \times [2^{nR_2}] \}$ of such generated codebooks, which constitutes an $(n,nR_1,nR_2)$ code for the DM-BC along with the optimal decoder. The random code ensemble generated in this manner is referred to as an $(n, nR_1,nR_2; p, \a, \e)$ \emph{Marton random code ensemble}, where $p=p(u_1,u_2)$ is the given pmf, $\a \in [0 \; 1]$ is the parameter used in step $(1)$, and $\e>0$ is the parameter used in steps $(1)$ and $(3)$. A rate pair $(R_1,R_2)$ is said to be \emph{achievable by the $(p, \a, \e)$-distributed Marton random code ensemble} if there exits a sequence of $(n, nR_1,nR_2; p, \a, \e)$ Marton random code ensembles such that
\[ 
\lim_{n \to \infty} \E_{\Cc_n} [ P_e^{(n)} (\Cc_n)] = 0,
\]
where the expectation is with respect to the Marton random codebook $\Cc_n$. Given $(p, \a, \e)$, let $\Rrbc^*(p, \a, \e)$ be the set of all rate pairs achievable by the $(p, \a, \e)$-distributed Marton random code ensemble. Given pmf $p=p(u_1,u_2)$ and function $x(u_1,u_2)$, the optimal rate region $\Rrbc^*(p)$, when it exists, is defined as
\[
\Rrbc^*(p) := \cl \left[ \bigcup_{\a \in [0 \; 1]} \lim_{\e \to 0} \Rrbc^*(p, \a, \e) \right].
\]

We are now ready to state main result of this section.
\begin{theorem}
\label{thm:opt_marton}
Given a pmf $p(u_1,u_2)$ and a function $x(u_1,u_2)$, the optimal rate region $\Rrbc^*(p)$ for the broadcast channel $p(y_1,y_2|x)$ is the closure of the set of rate pairs $(R_1,R_2)$ satisfying
\begin{subequations}
\label{eq:marton}
\begin{align}
\label{eq:thm_marton1}
R_1 &\le I(U_1;Y_1,U_2) - \a I(U_1;U_2), \\
\label{eq:thm_marton2}
R_1 &\le I(U_1,U_2;Y_1) - \min\{R_2;I(U_2;Y_1,U_1)-\ab I(U_1;U_2),I(U_1,U_2;Y_1)  \}, \\
R_2 &\le I(U_2;Y_2,U_1) - \ab I(U_1;U_2), \\
R_2 &\le I(U_1,U_2;Y_2) - \min\{R_1;I(U_1;Y_2,U_2)-\a I(U_1;U_2),I(U_1,U_2;Y_2)  \},
\end{align}
\end{subequations}%
for some $\a \in [0 \; 1]$.
\end{theorem}

We prove Theorem \ref{thm:opt_marton} by showing that given a pmf $p(u_1,u_2)$, a function $x(u_1,u_2)$, and $\a \in [0 \; 1]$, the rate region $\Rrbc^*(p,\a):=\cl\left[\lim_{\e \to 0} \Rrbc^*(p, \a, \e)  \right]$ is equal to the rate region characterized by (\ref{eq:marton}), which we will denote as $\Rrbc^{**}(p,\a)$. We take a two-step approach similar to Sections \ref{sec:achiev} and \ref{sec:converse}, and establish the achievability and the converse on the rate region $\Rrbc^*(p,\a)$, respectively.

The achievability proof is relegated to Appendix~\ref{app:marton_achiev}.
For the converse, given a fixed pmf $p=p(u_1,u_2)$, $\a \in [0 \; 1]$, and $\e>0$, we define the rate region $\Rrbc^{**}(p,\a,\d)$ as the set of rate pairs $(R_1,R_2)$ such that
\begin{subequations}
\begin{align}
\label{eq:marton_type1}
R_1 &\le I(U_1; Y_1,U_2) - \a I(U_1;U_2) + \d,\\
R_1 &\le I(U_1,U_2;Y_1) - \min\{R_2;I(U_2;Y_1,U_1)-\ab I(U_1;U_2),I(U_1,U_2;Y_1)  \} + \d,
\label{eq:marton_type2} \\ 
\label{eq:marton_type1_r2}
R_2 &\le I(U_2; Y_2,U_1) - \ab I(U_1;U_2) + \d,\\ 
\label{eq:marton_type2_r2}
R_2 &\le I(U_1,U_2;Y_2) - \min\{R_1;I(U_1;Y_2,U_2)-\a I(U_1;U_2),I(U_1,U_2;Y_2)  \} + \d.
\end{align}
\end{subequations}%
Note that the region $\Rrbc^{**}(p,\a,\d=0)$ is equal to $\Rrbc^{**}(p,\a)$ as defined in (\ref{eq:marton}). 

\begin{proposition}
\label{prop:outer_marton}
Let $p=p(u_1,u_2)$ be a pmf, $x(u_1,u_2)$ be a function, $\a \in [0 \; 1]$, and $\e>0$. If a rate pair $(R_1, R_2)$ is achievable by the $(p, \a, \e)$-distributed Marton random code ensemble, then there exists a continuous $\d'(\e)$ that tends to zero monotonically as $\e \to 0$ such that
\begin{equation}
\label{eq:outer_marton}
(R_1,R_2) \in \Rrbc^{**}(p,\a, \d'(\e)).
\end{equation}
In particular,
\begin{equation}
\label{eq:outer_marton_limit}
\Rrbc^{*}(p,\a) \sbq \Rrbc^{**}(p,\a).
\end{equation}
\end{proposition}

\begin{IEEEproof}
We first start with an averaged version of Fano's inequality for a Marton random code ensemble $\Cc_n$. Consider a fixed codebook $\Cc_n = \cc_n$. By Fano's inequality, 
\[
H(M_j | Y_j^n, \Cc_n= \cc_n) \le 1 + nR_j P_e^{(n)}(\cc_n)\quad j=1,2. 
\]
Taking the expectation over Marton random codebook $\Cc_n$, it follows that
\begin{equation}
\label{eq:marton_fano}
H(M_j | Y_j^n, \Cc_n) \le 1 + nR_j \E_{\Cc_n}[P_e^{(n)}(\Cc_n)] \le n\e_n, \quad j=1,2
\end{equation}%
for some $\e_n \to 0$ as $n \to \infty$ since $\E_{\Cc_n}[P_e^{(n)}(\Cc_n)] \to 0$.

We next define the indicator random variable
\begin{equation}
\label{eq:indicator_en_marton}
\Et_n = \mathbbm{1}_{\{ (U_1^n(M_1,L_1),U_2^n(M_2,L_2)) \in \aep(X_1,X_2) \}}.
\end{equation}%
Since $\Rh_1+ \Rh_2  = I(U_1;U_2) + 10 \e H(U_1,U_2)$, $\P(\Et_n = 0)$ tends to zero as $n \to \infty$ by the mutual covering lemma in~\cite[p.~208]{El-Gamal--Kim2011}. 

We are now ready to establish (\ref{eq:marton_type1}). For $n$ sufficiently large, we have
\begin{align} \notag
nR_1 &= H(M_1|M_2, \Cc_n) 
\\ \notag
& \overset{(a)}{\le} I(M_1;Y_1^n| M_2, \Cc_n) + n \e_n 
\\ \notag
& \le I(M_1, \Et_n ;Y_1^n| M_2, \Cc_n) + n \e_n 
\\ \notag
& \overset{(b)}{\le} 1 + I(M_1 ;Y_1^n| M_2, \Cc_n, \Et_n) + n \e_n 
\\ \notag 
& \le 1 + I(M_1 ;Y_1^n| M_2, \Cc_n, \Et_n = 0) \P(\Et_n = 0) + I(M_1 ;Y_1^n| M_2, \Cc_n, \Et_n = 1) \P(\Et_n = 1) + n \e_n 
\\ \notag
& \le 1 + nR_1 \P(\Et_n = 0) + I(M_1 ;Y_1^n| M_2, \Cc_n, \Et_n = 1) + n \e_n 
\\ \notag
& \le 1 + nR_1 \P(\Et_n = 0) + I(M_1,L_2 ;Y_1^n| M_2, \Cc_n, \Et_n = 1) + n \e_n 
\\ \notag
& \le 1 + nR_1 \P(\Et_n = 0) + n\Rh_2 + I(M_1;Y_1^n| M_2,L_2, \Cc_n, \Et_n = 1) + n \e_n 
\\ \notag
& = 1 + nR_1 \P(\Et_n = 0) + n\Rh_2 + \sum_{i=1}^{n} I(M_1;Y_{1i} | Y_1^{i-1}, M_2,L_2, \Cc_n, U_{2i}, \Et_n = 1) + n \e_n 
\\ \notag
& \le 1 + nR_1 \P(\Et_n = 0) + n\Rh_2 + \sum_{i=1}^{n} I(M_1, U_{1i}, Y_1^{i-1}, M_2,L_2, \Cc_n ;Y_{1i} |  U_{2i}, \Et_n = 1) + n \e_n 
\\ \notag
&\overset{(c)}{=} 1 + nR_1 \P(\Et_n = 0) + n\Rh_2 +\sum_{i=1}^{n} I( U_{1i} ;Y_{1i} |  U_{2i}, \Et_n = 1) + n \e_n
\\ \notag
& \overset{(d)}{\le} 1 + nR_1 \P(\Et_n = 0) + n\Rh_2 + n ( I( U_1 ;Y_1 |  U_2) + \d_2(\e)) + n \e_n,
\\ \notag
&\le 1 + nR_1 \P(\Et_n = 0) + n \ab (I(U_1;U_2)+ \d_1(\e) ) + n ( I( U_1 ;Y_1 |  U_2) + \d_2(\e)) + n \e_n,
\\ 
\label{eq:outer1_marton}
& \overset{(e)}{\le} n ( I( U_1 ;Y_1 ,  U_2) - \a I(U_1;U_2) + \d_3(\e)) + 2 n \e_n,
\end{align}%
where $(a)$ follows by (the averaged version of) Fano's inequality in (\ref{eq:marton_fano}), $(b)$ follows since $\Et_n$ is a binary random variable, $(c)$ follows since $(M_1,M_2, Y_1^{i-1}, \Cc_n, \Et_n) \to (U_{1i}, U_{2i})\to Y_{1i}$ form a Markov chain for every $i \in [n]$, $(d)$ follows by the memoryless property of the channel and by Lemma~\ref{lem:uniform_tyep_iid} in Appendix~\ref{app:lemma_pmf} since the distribution of $(U_1^n(M_1,L_1), U_2^n(M_2,L_2))$ is permutation invariant by construction, and $(e)$ follows since $\P(\Et_n = 0)$ tends to zero as $n \to \infty$. 

For the proof of (\ref{eq:marton_type2}), we start with
\begin{align} \notag
nR_1 &= H(M_1|M_{2}, \Cc_n) 
\\ \notag
& \overset{(a)}{\le} I(M_1;Y_1^n| M_2, \Cc_n) + n \e_n 
\\
\label{eq:outer2_0_marton}
& = I(M_1, M_2; Y_1^n|\Cc_n) - I(M_2; Y_1^n |\Cc_n) + n \e_n,
\end{align}%
where $(a)$ follows by (the averaged version of) Fano's inequality in (\ref{eq:marton_fano}). Following arguments similar to (\ref{eq:outer1_marton}), the first term in (\ref{eq:outer2_0_marton}) can be bounded as
\begin{align} \notag
I(M_1, M_2; Y_1^n|\Cc_n) &\le 1 + n(R_1+R_2) \P(\Et_n = 0) + \sum_{i=1}^n I(M_1, M_2; Y_{1i} | \Cc_n, Y_1^{i-1}, \Et_n=1)
\\ \notag
& \le n \e_n + \sum_{i=1}^n I(M_1, M_2, \Cc_n, Y_1^{i-1} ; Y_{1i} | \Et_n = 1) 
\\ \notag
&= n \e_n + \sum_{i=1}^n I(M_1, M_2, \Cc_n, Y_1^{i-1}, U_{1i}, U_{2i} ; Y_{1i} | \Et_n = 1) 
\\ \notag
&= n \e_n + \sum_{i=1}^n I(U_{1i}, U_{2i} ; Y_{1i} | \Et_n = 1),
\\
\label{eq:outer2_1_marton}
&\le n \e_n + n (I(U_1, U_2 ; Y_1) + \d_5(\e)).
\end{align}

For the second term in (\ref{eq:outer2_0_marton}), we need the following lemma, which is proved in Appendix~\ref{app:list_marton}. This lemma is a version of Lemma \ref{lem:outer2_list} for Marton random code ensembles.

\begin{lemma}
\label{lem:outer2_list_marton} 
For every $\e' > \e$ and for $n$ sufficiently large,
\[
I(M_2; Y_1^n |\Cc_n) \ge n [ \min \{ R_2, I(U_2;Y_1,U_1) - \ab I(U_1;U_2), I(U_1,U_2;Y_1) \} - \d_6(\e') ]- n \e_n.
\]
\end{lemma}

Combining (\ref{eq:outer2_0_marton}), (\ref{eq:outer2_1_marton}), and Lemma \ref{lem:outer2_list_marton} with $\e' = 2 \e$, we have
\begin{align}
nR_1 \le n [I(U_1,U_2;Y_1) - \min \{ R_2, I(U_2;Y_1,U_1)-\ab I(U_1;U_2), I(U_1,U_2;Y_1) \} + \d_7(\e)] + 2 n \e_n
\label{eq:outer2_marton}
\end{align}
for $n$ sufficiently large. 

For (\ref{eq:marton_type1_r2}) and (\ref{eq:marton_type2_r2}), we can similarly establish for receiver $2$
\begin{equation}
\label{eq:outer1_marton2}
nR_2 \le n ( I( U_2 ;Y_2 ,  U_1) - \ab I(U_1;U_2) + \d_4(\e)) + 2 n \e_n
\end{equation}
and
\begin{align}
nR_2 \le n [I(U_1,U_2;Y_2) - \min \{ R_1, I(U_1;Y_2,U_2)-\a I(U_1;U_2), I(U_1,U_2;Y_2) \} + \d_8(\e)] + 2 n \e_n
\label{eq:outer2_marton2}
\end{align}
for $n$ sufficiently large. The proof of (\ref{eq:outer_marton}) follows by letting $n \to \infty$ in (\ref{eq:outer1_marton}), (\ref{eq:outer2_marton}), (\ref{eq:outer1_marton2}), and (\ref{eq:outer2_marton2}) and taking a continuous monotonic function $\d'(\e) \ge \max \{\d_3(\e), \d_4(\e), \d_7(\e), \d_8(\e) \}$ that tends to zero as $\e \to 0$. Letting $\e \to 0$ in (\ref{eq:outer_marton}) establishes (\ref{eq:outer_marton_limit}), which completes the proof of Proposition~\ref{prop:outer_marton}.
\end{IEEEproof}

\begin{remark}
Marton coding we have analyzed involves two codewords. Marton's original coding scheme~\cite{Marton1979} uses rate splitting and superposition coding, and involves an additional codeword that carries messages for both receivers (see also~\cite[Proposition 8.1]{El-Gamal--Kim2011}). Our technique can be similarly adapted to this general version of Marton coding.
\end{remark}

\section{Discussion}
\label{sec:discuss}
For the linear computation problem, the outer bound on the optimal rate region presented in Section~\ref{sec:converse} is valid for \emph{any} computation, not only for natural computation. The inner bound presented in Theorem~\ref{thm:achiev}, however, matches with this outer bound only for \emph{natural} computation. It is an interesting but difficult problem to characterize the optimal rate region for an arbitrary linear computation problem. At this point, it is unclear whether it is the inner or the outer bound that is loose. The extension of the results in this paper to more than two senders is also a challenging question. 

A more fundamental question is to establish a general outer bound on the \emph{capacity region} of the linear computation problem. When $(X_1,X_2)\to W_{\av} \to Y$ form a Markov chain and $a_1,a_2 \neq 0$, we can establish the following outer bound by using Fano's inequality. If a rate pair $(R_1, R_2)$ is achievable, then
\begin{subequations}
\label{eq:general_outer_comp}
\begin{align}
R_1 &\le I(X_1; Y|X_2, Q),\\
R_2 &\le I(X_2; Y|X_1, Q),\\
R_1 &\le I(W_{\av}; Y|Q)-I(X_2; W_{\av}|T, Q),\\
R_2 &\le I(W_{\av}; Y|Q)-I(X_1; W_{\av}|T, Q),\\
R_1+R_2 &\le I(W_{\av}; Y|Q) + I(X_1, X_2; W_{\av}|T, Q) - I(X_1; W_{\av}|T, Q)-I(X_2; W_{\av}|T, Q)
\end{align}
\end{subequations}%
for some $p(q)p(x_1|q)p(x_2|q)p(t|x_1,x_2, q)$ such that $W_{\av}\to(X_1, X_2)\to T$. Suppose that we set $Q = \emptyset$ and fix a pmf $p=p(x_1)p(x_2)$ in (\ref{eq:general_outer_comp}). If the auxiliary random variable $T = \emptyset$, (\ref{eq:general_outer_comp}) reduces to the rate region $\Rrc(p)$ in Section~\ref{sec:main_result}. If $T = (X_1,X_2)$, (\ref{eq:general_outer_comp}) reduces to the rate region $\Rrmac(p)$ in Section~\ref{sec:main_result}. Thus, we can conclude that this general outer bound recovers as extreme special cases the components of the outer bound in Theorem~\ref{thm:outer_ncc} that was established for a random ensemble of homologous codes. Whether and when both outer bounds coincide after taking time sharing and the union over all $p$ is left as another open problem.

%\section*{Acknowledgments}
%The work in this paper was supported in part by the Electronics and Telecommunications Research Institute
%through Grant 17ZF1100 from the Korean Ministry of Science, ICT, and Future Planning.

%%%%%%% APPENDIX %%%%%%%%%%
%\newpage
\appendices
\section{Proof of Proposition~\ref{prop:opt_reg}}
\label{app:prop1}
Fix pmf $p = p(x_1)p(x_2)$. We first show that $[\Rr_{CF}(p) \cup \Rrmac(p) ] \sbq \Rr^*(p)$.
Suppose that the rate pair $(R_1, R_2) \in \Rr_{CF}(p)$. Then, for every $j \in \{1,2\}$ such that $a_j \neq 0$, the rate pair $(R_1,R_2)$ satisfies
\begin{align*}
R_j &\le H(X_j) - H(W_{\av} |Y) \\
&\le H(X_j) - H(W_{\av} |Y, X_{j^c}) \\
&= I(X_j;Y|X_{j^c}),
\end{align*}
and
\begin{align*}
R_j &\le H(X_j) - H(W_{\av} |Y) \\
&= I(X_1,X_2;Y) - I(X_{j^c};W_{\av},Y) \\
&\le I(X_1,X_2;Y) - \min\{R_{j^c},I(X_{j^c};W_{\av},Y)\},
\end{align*}
which implies that $(R_1,R_2) \in \Rr^*(p)$. It follows that $\Rr_{CF}(p) \sbq \Rr^*(p)$. Similarly, suppose that the rate pair $(R_1,R_2) \in \Rrmac(p)$. Then, for every $j \in \{1,2\}$ such that $a_j \neq 0$, the rate pair $(R_1,R_2)$ satisfies
\[
R_j \le I(X_j;Y|X_{j^c}),
\]
and
\begin{align*}
R_j &\le I(X_1,X_2;Y) - R_{j^c} \\
& \le I(X_1,X_2;Y) - \min \{R_{j^c}, I(X_{j^c};W_{\av}, Y)\},
\end{align*}
which implies that $(R_1,R_2) \in \Rr^*(p)$. Therefore, $\Rrmac(p) \sbq \Rr^*(p)$.

Next, we show that $\Rr^*(p) \sbq [\Rr_{CF}(p) \cup \Rrmac(p)]$.
Suppose that the rate pair $(R_1,R_2) \in \Rr^*(p)$ such that $R_{j^c} > I(X_{j^c};W_{\av},Y)$ for each $j \in \{1,2\}$ with $a_j \neq 0$. Then, $(R_1,R_2)$ satisfies
\begin{align*}
R_j &\le I(X_1,X_2;Y) - I(X_{j^c};W_{\av},Y) \\
&= H(X_j) - H(W_{\av}|Y),
\end{align*}
for each $j \in \{1,2\}$ with $a_j \neq 0$. Then, $(R_1,R_2) \in \Rr_{CF}(p)$. It is easy to see that the rate pair $(R_1,R_2) \in \Rr^*(p)$ that satisfies $R_{j^c} \le I(X_{j^c};W_{\av},Y)$ for one or more $j \in \{1,2\}$ with $a_j \neq 0$, is included in $\Rrmac(p)$. Thus, $\Rr^*(p) \sbq [\Rr_{CF}(p) \cup \Rrmac(p)]$, which completes the proof.

\section{}
\label{app:full_rank}
\begin{lemma}
\label{lem:full_rank_G}
Let $G$ be an $ nR \times n$ random matrix over $\Fq$ with $R < 1$ where each element is drawn i.i.d. $\U(\Fq)$. Then,
\[
\lim_{n \to \infty} n \P(G \textrm{ is not full rank}) = 0.
\]
\end{lemma}

\begin{IEEEproof}
Probability of choosing $ nR $ linearly independent rows can be written as
\begin{align*}
\P(G \textrm{ is full rank}) &= \frac{\prod_{j=1}^{ nR } (q^n-q^{j-1})}{(q^n)^{ nR }} 
\\
& = \prod_{j=1}^{ nR } (1-q^{j-1-n}) \\
&\ge (1-q^{-n(1-R)})^{nR}.
\end{align*}
Using this relation, we have
\begin{align*}
n \P(G \textrm{ is not full rank}) &= n (1 - \P(G \textrm{ is full rank}) )
\\
& \le n (1 - (1-q^{-n(1-R)})^{nR})
\\
& \overset{(a)}{\le} n^2 R q^{-n(1-R)},
\end{align*}
where $(a)$ follows by Bernoulli's inequality. Since $R < 1$, $\lim_{n \to \infty} n^2 q^{-n(1-R)} = 0$, which completes the proof.
\end{IEEEproof}

\section{Proof of Lemma~\ref{lem:equiv_reg}}
\label{app:lemma1}
Fix pmf $p = p(x_1)p(x_2)$. We will show that if the condition in (\ref{eq:comp_favor}) holds, then $\Rr_{CF}(p) \cup \Rr_1(p) \cup \Rr_2(p)  = \Rr_{CF}(p) \cup \Rrmac(p)$. By definition of the rate regions $ \Rr_1(p), \Rr_2(p)$ and $\Rrmac(p)$, it is easy to see that $\Rr_{CF}(p) \cup \Rr_1(p) \cup \Rr_2(p)  \sbq \Rr_{CF}(p) \cup \Rrmac(p)$
holds in general. Then, it suffices to show that if the condition in (\ref{eq:comp_favor}) holds, then $\Rrmac(p) \sbq [\Rr_{CF}(p) \cup \Rr_1(p) \cup \Rr_2(p)]$. Suppose that the condition in (\ref{eq:comp_favor}) is satisfied. Let the rate pair $(R_1,R_2) \in \Rrmac(p)$ be such that $R_{j^c} > I(X_{j^c};W_{\av},Y)$ for every $j \in \{1,2\}$ with $a_j \neq 0$. Then, $(R_1,R_2)$ satisfies
\begin{align*}
R_j &\le I(X_1,X_2;Y) - I(X_{j^c};W_{\av},Y)  \\
&= H(X_j) - H(W_{\av}|Y),
\end{align*}
for each $j \in \{1,2\}$ with $a_j \neq 0$, implying that $(R_1,R_2) \in \Rr_{CF}(p)$. Now, let the rate pair $(R_1,R_2) \in \Rrmac(p)$ be such that $R_{j^c} \le I(X_{j^c};W_{\av},Y)$ for some $j \in \{1,2\}$ with $a_j \neq 0$. By condition (\ref{eq:comp_favor}), we have 
\begin{align*}
I(X_{j^c};W_{\av},Y) &= I(X_1,X_2;Y) - H(X_j) + H(W_{\av}|Y) \\
&= I(X_1,X_2;Y) - H(X_j) + \min_{\bv \neq 0} H(W_{\bv}|Y) \\
&\le I(X_1,X_2;Y) - H(X_j) + \min_{\bv \in \Fqh^{1 \times 2}} H(W_{\bv}|Y).
\end{align*}
Then, the rate pair $(R_1,R_2) \in \Rr_1(p) \cup \Rr_2(p)$, which completes the proof.

\section{Proof of Lemma~\ref{lem:Fano_homolog}}
\label{app:fano_homolog}
Suppose that $a_j \neq 0$. Then, 
\begin{equation}
\label{eq:fano_proof1}
H(M_j|Y^n,M_{j^c},\Cc_n) = I(M_j; W^n_{\av} | Y^n, M_{j^c}, \Cc_n) + H(M_j|W^n_{\av},Y^n,M_{j^c},\Cc_n).
\end{equation}
To bound the first term in (\ref{eq:fano_proof1}), we need a version of Fano's inequality for computation.
\begin{lemma}
\label{lem:Fano_compute}
If the average probability of error $\E_{\Cc_n} [ P_e^{(n)}(\Cc_n)]$ tends to zero as $n \to \infty$, then
\[
H(W_{\av}^n|Y^n, \Cc_n) \le n \e_n
\] 
for some $\e_n \to 0$ as $n \to \infty$.
\end{lemma}

\begin{IEEEproof}
For fixed codebook $\Cc_n = \cc_n$, by Fano's inequality
\[
H(W_{\av}^n|Y^n, \Cc_n = \cc_n) \le 1 + n P_e^{(n)}(\cc_n). 
\]
Taking the expectation over the random homologous codebook $\Cc_n$, we have
\[
H(W_{\av}^n|Y^n, \Cc_n ) \le 1 + n E_{\Cc_n} [ P_e^{(n)}(\Cc_n)] \overset{(a)}{\le} n \e_n,
\]
where $(a)$ follows since $\E_{\Cc_n} [ P_e^{(n)}(\Cc_n)]$ tends to zero as $n \to \infty$.
\end{IEEEproof}

Combining (\ref{eq:fano_proof1}) with Lemma~\ref{lem:Fano_compute}, we have
\begin{align*}
H(M_j|Y^n,M_{j^c},\Cc_n) &\le n \e_n + H(M_j|W^n_{\av},Y^n,M_{j^c},\Cc_n) 
\\
& \overset{(a)}{=} n \e_n + H(M_j|W^n_{\av}, X_{j^c}^n(M_{j^c}), Y^n,M_{j^c},\Cc_n)
\\
& \overset{(b)}{=} n \e_n + H(M_j|W^n_{\av}, X_{j}^n(M_j), X_{j^c}^n(M_{j^c}), Y^n,M_{j^c},\Cc_n)
\\
& \le n \e_n + H(M_j| X_{j}^n(M_j), \Cc_n)
\\
& \overset{(d)}{\le} 2 n\e_n
\end{align*}
where $(a)$ follows since $X^n_{j^c}(M_{j^c})$ is a function of $(M_{j^c}, \Cc_n)$, $(b)$ follows since $a_j \neq 0$ and $X^n_j(M_j)$ is a function of $(X^n_{j^c}(M_{j^c}), W^n_{\av})$, and $(d)$ follows since $H(M_j | X_j^n(M_j), \Cc_n)$ tends to zero as $n \to \infty$.

\section{Proof of Lemma~\ref{lem:output_ncc_input}}
\label{app:lemma_pmf}
Let $i \in [n]$, and $(x,y) \in \Fq \times \Yc$. Then,
\begin{align} \notag
\P(X_i = x, Y_i = y | X^n \in \aep(X)) &= \P(X_i = x| X^n \in \aep(X)) \P(Y_i = y | X_i = x, X^n \in \aep(X)) \\
&= \P(X_i = x| X^n \in \aep(X)) p_{Y|X}(y | x). 
\label{eq:dist_proof}
\end{align}
We make a connection between the conditional distribution of $X_i$ given $\{X^n \in \aep(X)\}$ and the input pmf $p(x)$. Therefore, we start with exploring the conditional distribution of $X_i$ given $\{X^n \in \aep(X)\}$.

\begin{lemma}
\label{lem:ncc_uniform_type}
Let $p_X$ be a pmf on $\Fq$, and $\e > 0$. Define $\aep(X,\Th)$ as the set of elements in $\aep(X)$ with type $\Th$. Suppose $X^n(m) = U^n(m,L(m))$ denote the random codeword assigned to message $m$ by $(n,nR;p_X,\e)$ random homologous code ensemble. Then,
\[
U^n(m,L) | \{ U^n(m,L) \in \aep(X,\Th) \} \sim \U(\aep(X,\Th)),
\]
for every $m \in \Fq^{nR}$.
\end{lemma}

\begin{IEEEproof}
Without loss of generality, we drop index $m$. It suffices to show that the distribution of $U^n(L)$ is permutation invariant. Let $u^n, v^n$ have the same type (typical or not) and let $u^n = \sigma(v^n)$ for some permutation $\sigma$. Then, we have
\begin{align*}
\P(U^n(L)=u^n ) &= \sum_{l} \sum_{\Gb} \P( L=l, G = \Gb, D^n = u^n \ominus l \Gb) 
\\
&\overset{(a)}{=} \sum_{l} \sum_{\Gb} \P( L=l, G = \sigma(\Gb), D^n = v^n \ominus l \sigma(\Gb)) 
\\
&= \P(U^n(L)=v^n ),
\end{align*}
where $\sigma(\Gb)$ is the matrix constructed by applying permutation $\sigma$ to the columns of $\Gb$, and $(a)$ follows since a permutation applied to a coset code preserves the type of each codeword.
\end{IEEEproof}

Building on top of Lemma~\ref{lem:ncc_uniform_type}, we next establish that the conditional distribution of $X_i$ given $\{X^n \in \aep(X)\}$ is \emph{close} to the input pmf $p(x)$. 

\begin{lemma}
\label{lem:uniform_tyep_iid}
Let $\e > 0$. Define $\aep(X,\Th)$ in a similar way to Lemma~\ref{lem:ncc_uniform_type}. Suppose that the distribution of $X^n$ is uniform within each type in the typical set, namely, for each type $\Th$
\begin{equation}
\label{eq:cond_lem_iid}
X^n | \{ X^n \in \aep(X,\Th) \} \sim \U(\aep(X,\Th)).
\end{equation}

Then, conditional on the typical set, $X_i$'s have identical distribution that satisfies
\[
(1-\e) p(x) \le P(X_i = x | X^n \in \aep(X)) \le (1+\e) p(x), \quad \forall x \in \Xc.
\] 
\end{lemma}

\begin{IEEEproof}
Let $x \in \Xc$. For a type $\Th$, let $\Th_x$ denote the empirical mode of $x$ within type $\Th$. Then, for every type $\Th$ within the set $\aep(X)$, we have
\begin{align*}
\P(X_i = x | X^n \in \aep(X,\Th)) &= \sum_{x^n \in \aep(X,\Th) \atop \textrm{s.t. } x_i = x} \P(X^n = x^n | X^n \in \aep(X,\Th)) \\
&\overset{(a)}{=} \sum_{x^n \in \aep(X,\Th) \atop x_i = x} \frac{1}{| \aep(X,\Th) |} \\
&\overset{(b)}{=}   \Th_x | \aep(X,\Th) | \frac{1}{| \aep(X,\Th) |} \\
&= \Th_x,
\end{align*}
where $(a)$ follows since $X^n$ is conditionally uniform over $\aep(X,\Th)$, and $(b)$ follows since $\aep(X,\Th)$ is closed under permutation. Combining this observation with the fact that $\Th$ is the type of a typical sequence, we get
\[
(1-\e) p(x) \le \P(X_i = x | X^n \in \aep(X,\Th)) \le (1+\e) p(x), \quad \forall x \in \Xc.
\]
Since $\aep(X)$ is the disjoint union of $\aep(X,\Th)$ over all types, multiplying each side with $\P(X^n \in \aep(X,\Th))$ and then summing over $\Th$ gives
\[
(1-\e) p(x) \P(X^n \in \aep(X)) 
\le \P(X_i = x , X^n \in \aep(X)) 
\le (1+\e) p(x) \P(X^n \in \aep(X)),
\]
for all $x \in \Xc$. The claim follows from dividing each side by $\P(X^n \in \aep(X))$.
\end{IEEEproof}

Back to the proof of Lemma~\ref{lem:output_ncc_input}, we have by Lemma \ref{lem:ncc_uniform_type} that the distribution of $X^n$ satisfies the condition in (\ref{eq:cond_lem_iid}) in Lemma~\ref{lem:uniform_tyep_iid}. Therefore, combining (\ref{eq:dist_proof}) with Lemma~\ref{lem:uniform_tyep_iid} completes the proof.

\section{Proof of Lemma~\ref{lem:outer2_list}}
\label{app:list_homolog}
Let $\e'' > \e'$. Suppose that $a_j \neq 0$, and $j^c = \{1,2\} \setminus \{j\}$. First, by Lemma~\ref{lem:Fano_compute}, we have
\[
I(M_{j^c}; Y^n |\Cc_n) \ge  I(M_{j^c}; W_{\av}^n, Y^n |\Cc_n) - n \e_n.
\]
Therefore, it suffices to prove that for $n$ sufficiently large,
\[
I(M_{j^c}; W_{\av}^n, Y^n |\Cc_n) \ge n [\min \{ R_{j^c}, I(X_{j^c};W_{\av}, Y) \} - \d_4(\e'') - \e_n].
\]
Similar to~\cite{Bandemer--El-Gamal--Kim2012a}, we will show that given $W_{\av}^n, Y^n$, and $\Cc_n$, a relatively short list $\Lc \sbq \Fq^{nR_{j^c}}$ can be constructed that contains $M_{j^c}$ with high probability. Define a random set
\begin{align*}
\Lc = \{ m \in \Fq^{nR_{j^c}}: (X_{j^c}^n(m), W_{\av}^n, Y^n) \in \aepvarvar(X_{j^c},W_{\av}, Y) \}.
\end{align*}
Define two events $\Mc_1 = \{M_1 = M_2 = \zerov \}$ and $\Mc_2 = \{L_1(M_1) = L_2(M_2) = \zerov\}$. The indicator random variable $E_n$ is as defined in (\ref{eq:indicator_en}). By the symmetry of the codebook generation, for each $m \in \Fq^{nR_{j^c}}, m \neq M_{j^c}$, we have 
\begin{align*}
\P( m \in \Lc,&\,  E_n = 1) 
\\
&= \P( m \in \Lc, E_n = 1 | \Mc_1, \Mc_2) 
\\
&=  \P( (X_{j^c}^n(m), W_{\av}^n, Y^n) \in \aepvarvar, (X_1^n(\zerov),X_2(\zerov)) \in \aepvar | \Mc_1, \Mc_2) 
\\
& \le  \P( (U_{j^c}^n(m,l), W_{\av}^n, Y^n) \in \aepvarvar \textrm{ for some } l \in \Fq^{n\Rh_{j^c}}, (X_1^n(\zerov),X_2(\zerov)) \in \aepvarvar | \Mc_1, \Mc_2)
\\
& \le \sum_{l} \sum_{(x_1^n,x_2^n) \in \atop \aepvarvar(X_1,X_2)} \sum_{(u^n,w^n,y^n) \in \atop \aepvarvar(X_{j^c},W_{\av},Y)} 
\P \left( \begin{array}{c | c}
 U_{j^c}^n(m,l) = u^n, a_1 D_1^n \oplus a_2 D_2^n = w^n, & \Mc_1, \\
 D_1^n = x_1^n, D_2^n = x_2^n, Y^n = y^n & \Mc_2
\end{array} \right)
\\
&= \sum_{l} \sum_{(x_1^n,x_2^n) \in \atop \aepvarvar(X_1,X_2)} \sum_{(u^n,w^n,y^n) \in \atop \aepvarvar(X_{j^c},W_{\av},Y)} 
\P \left( \begin{array}{c | c}
 U_{j^c}^n(m,l) = u^n,  \\
 a_1 D_1^n \oplus a_2 D_2^n = w^n, & \Mc_1, \Mc_2 \\
 D_1^n = x_1^n, D_2^n = x_2^n
\end{array} \right) p(y^n|x_1^n,x_2^n) 
\\
& \overset{(a)}{\le} q^{n(\Rh_1+ \Rh_2)} \sum_{l} \sum_{(x_1^n,x_2^n) \in \atop \aepvarvar(X_1,X_2)} \sum_{(u^n,w^n,y^n) \in \atop \aepvarvar(X_{j^c},W_{\av},Y)} 
\P \left( \begin{array}{c | c}
 U_{j^c}^n(m,l) = u^n,\\
  a_1 D_1^n \oplus a_2 D_2^n = w^n, \\
 D_1^n = x_1^n, D_2^n = x_2^n
\end{array} \, \Mc_1 \right) p(y^n|x_1^n,x_2^n) 
\\
&= q^{n(\Rh_1+ \Rh_2)} \sum_{l} \sum_{(x_1^n,x_2^n) \in \atop  \aepvarvar(X_1,X_2)} \sum_{(w^n,y^n) \in \atop \aepvarvar(W_{\av},Y)} \\
& \quad\quad\quad\quad\quad
 \sum_{u^n \in \atop \aepvarvar(X_{j^c}|w^n,y^n)} 
 \P \left( \begin{array}{c}
 [m \; l] G \oplus D_{j^c}^n = u^n, \\
 D_1^n = x_1^n, D_2^n = x_2^n
\end{array} \right) p(y^n|x_1^n,x_2^n) \,
\mathbbm{1}_{\{w^n = a_1 x_1^n \oplus a_2 x_2^n\}}
\\
&= q^{n(\Rh_1+ \Rh_2)} \sum_{l} \sum_{(x_1^n,x_2^n) \in \atop  \aepvarvar(X_1,X_2)} \sum_{(w^n,y^n) \in \atop \aepvarvar(W_{\av},Y)} 
 \sum_{u^n \in \atop \aepvarvar(X_{j^c}|w^n,y^n)} 
 q^{-3n} \, p(y^n|x_1^n,x_2^n) \,
\mathbbm{1}_{\{w^n = a_1 x_1^n \oplus a_2 x_2^n\}}
\\
& \le q^{n(\Rh_1+\Rh_2+\Rh_{j^c})} \:
q^{-3n} \:  q^{n(H(X_{j^c}|W_{\av},Y)+H(X_1)+H(X_2) +\d_4(\e''))}
\\
& \overset{(b)}{\le} q^{-n (I(X_{j^c};W_{\av},Y) - \d_4(\e'') - 3 \e )},
\\
& \le q^{-n (I(X_{j^c};W_{\av},Y) - \d_4(\e'') )},
\end{align*}%
where $(a)$ follows by~\cite[Lemma 11]{Lim--Gastpar2016}, and $(b)$ follows by the construction of the random homologous codebook $\Cc_n$ with $\Rh_i = D(p_{X_i} \| \U(\Fq)) + \e$. Since $\P(E_n = 1)$ tends to one as $n \to \infty$, for $n$ sufficiently large we have $\P(E_n=1) \ge q^{-\e}$. Therefore, for $n$ sufficiently large, the conditional probability is bounded as follows
\begin{align*}
\P( m \in \Lc | E_n = 1) &= \frac{\P( m \in \Lc , E_n = 1)}{\P(E_n=1)}
\\
& \le \P( m \in \Lc , E_n = 1) q^{\e}.
\end{align*}
The expected cardinality of $\Lc$ given  $\{E_n = 1 \}$ is then bounded as
\begin{align} \notag
\E(|\Lc| | E_n = 1) & \le 1 + \sum_{m \neq M_{j^c}} \P( m \in \Lc | E_n = 1)
\\
& \le 1+ q^{n(R_{j^c} - I(X_{j^c};W_{\av},Y) + \d_4(\e'') + \frac{\e}{n}) }
\\
&= 1+ q^{n(R_{j^c} - I(X_{j^c};W_{\av},Y) + \d_4(\e'') + \e_n) },
\label{eq:ent_list}
\end{align}
for $n$ sufficiently large. 
Define another indicator random variable $F_n = \mathbbm{1}_{\{ M_{j^c} \in \Lc \}}$. Since $\e'' > \e'$ and $\P(E_n=1)$ tends to one as $n \to \infty$, by the conditional typicality lemma in~\cite[p.~27]{El-Gamal--Kim2011}, $\P(F_n=1)$ tends to one as $n \to \infty$. Then, for $n$ sufficiently large, we have
\begin{align} \notag 
H(M_{j^c} | \Cc_n, & W_{\av}^n, Y^n) 
\\\notag
&= H(M_{j^c} | \Cc_n, W_{\av}^n, Y^n, E_n, F_n) + I ( M_{j^c}; E_n, F_n | \Cc_n, W_{\av}^n, Y^n)
\\ \notag
& \le H(M_{j^c} | \Cc_n, W_{\av}^n, Y^n, E_n, F_n) + 2 
\\ \notag
& \le 2 + \P(F_n=0) H(M_{j^c} | \Cc_n, W_{\av}^n, Y^n, F_n=0, E_n) + H(M_{j^c} | \Cc_n, W_{\av}^n, Y^n, F_n=1, E_n)
\\
& \le 2 +  nR_{j^c} \P(F_n=0) + H(M_{j^c} | \Cc_n, W_{\av}^n, Y^n, F_n=1, E_n).
\label{eq:list_comp}
\end{align}

For the last term in (\ref{eq:list_comp}), we use the fact that if $M_{j^c} \in \Lc$, then the conditional entropy cannot exceed $\log(|\Lc|)$:
{\allowdisplaybreaks
\begin{align*}
H(M_{j^c} | \Cc_n, W_{\av}^n, & Y^n, F_n=1, E_n) 
\\
& \overset{(a)}{=} H(M_{j^c} | \Cc_n, W_{\av}^n, Y^n, F_n=1, E_n, \Lc, |\Lc|) 
\\
& \le H(M_{j^c} | F_n=1, E_n, \Lc, |\Lc|) 
\\
&= \sum_{l=0}^{q^{nR_{j^c}}} \P(|\Lc|=l, E_n = 1) H(M_{j^c} | E_n=1, F_n = 1, \Lc, |\Lc| = l)  \\
& \quad +\sum_{l=0}^{q^{nR_{j^c}}}  \P(|\Lc|=l, E_n = 0) H(M_{j^c} | E_n=0, F_n = 1, \Lc, |\Lc| = l)
\\
& \le \sum_{l=0}^{q^{nR_{j^c}}}  \P(|\Lc|=l, E_n = 1) H(M_{j^c} | E_n=1, F_n = 1, \Lc, |\Lc| = l) + 
 \P(E_n = 0) nR_{j^c}
\\
& \le \sum_{l=0}^{q^{nR_{j^c}}}  \P(|\Lc|=l, E_n = 1) \log(l) +  nR_{j^c} \P(E_n = 0) 
\\
& \le \sum_{l=0}^{q^{nR_{j^c}}}  \P(|\Lc|=l | E_n = 1) \log(l) +  nR_{j^c}  \P(E_n = 0) 
\\
&= \E [\log(|\Lc|) | E_n = 1] + nR_{j^c} \P(E_n = 0)  
\\
& \overset{(b)}{\le} \log(\E [|\Lc| | E_n = 1]) + nR_{j^c} \P(E_n = 0)  
\\
& \overset{(c)}{\le} 1 + \max \{0, n(R_{j^c} - I(X_{j^c};W_{\av},Y) + \d_4(\e'') + \e_n) \} + nR_{j^c} \P(E_n = 0)
\\
& \le 1 + \max \{0, n(R_{j^c} - I(X_{j^c};W_{\av},Y)) \} + n\d_4(\e'') + n\e_n + nR_{j^c} \P(E_n = 0)
\end{align*}
}%
where $(a)$ follows since the set $\Lc$ and its cardinality $| \Lc|$ are functions of $(\Cc_n, W_{\av}^n,Y^n)$, $(b)$ follows by Jensen's inequality, and $(c)$ follows by (\ref{eq:ent_list}) and the soft-max interpretation of the log-sum-exp function~\cite[p.~72]{Boyd--Vandenberghe2004}.
Substituting back gives
\begin{align*}
I(M_{j^c}; W_{\av}^n, Y^n | \Cc_n) &= H(M_{j^c} | \Cc_n) - H(M_{j^c} | \Cc_n, W_{\av}^n, Y^n)
\\
&= nR_{j^c} - H(M_{j^c} | \Cc_n, W_{\av}^n, Y^n)
\\
& \ge nR_{j^c} - 2 -  nR_{j^c} \P(F_n=0) - H(M_{j^c} | \Cc_n, W_{\av}^n, Y^n, F_n=1, E_n)
\\
& \ge nR_{j^c} - 3 - nR_{j^c} (\P(E_n=0) + \P(F_n=0))\\
&\quad\quad\quad
- \max \{0, n(R_{j^c} - I(X_{j^c};W_{\av},Y)) \} - n\d_4(\e'') - n\e_n
\\
&=  n [ \min \{ R_{j^c}, I(X_{j^c};W_{\av},Y) \} - \d_4(\e'') - \e_n ] - 3 - nR_{j^c} (\P(E=0) + \P(F = 0)) 
\\
&\overset{(a)}{=} n [ \min \{ R_{j^c}, I(X_{j^c};W_{\av},Y)\} - \d_4(\e'') - 2 \e_n],
\end{align*}
where $(a)$ follows for large $n$ since both probabilities $\P(E_n=0)$ and $\P(F_n=0)$ tend to zero as $n \to \infty$.

\section{Proof of Achievability for Theorem~\ref{thm:opt_marton}}
\label{app:marton_achiev}
Let $\a \in [0 \; 1]$ and $\e>0$. Consider an $(n,nR_1,nR_2;p,\a,\e)$ Marton random code ensemble. We use the nonunique simultaneous joint typicality decoding rule in~\cite{Wang--Sasoglu--Bandemer--Kim2013} to establish the achievability. Let $\e' > \e$. Upon receiving $y_j^n$ at receiver $j=1,2$, the $\e'$-joint typicality decoder $j$ looks for a unique $m_j \in [2^{nR_j}]$ such that
\[
(u_1^n(m_1,l_1), u_2^n(m_2,l_2),y_j^n) \in \aepvar(U_1,U_2,Y_j),
\]
for some $l_1 \in [2^{n\Rh_1}]$, $l_2 \in [2^{n\Rh_2}]$ and $m_{j^c} \in [2^{n R_{j^c}}]$, where $j^c$ denotes $\{1,2\} \setminus j$. If the decoder $j=1,2$ finds such $m_j$, then it declares $m_j$ as an estimate; otherwise, it declares an error.

We analyze the probability of error. It suffices to consider decoder 1, which declares an error if one or more of the following events occur
\begin{align*}
\Ec_0 &= \{ (U_1^n(M_1,l_1),U_2^n(M_2,l_2)) \notin \aep(U_1,U_2) \textrm{ for every } (l_1,l_2) \in [2^{n\Rh_1}] \times [2^{n\Rh_2}]\}, \\
\Ec_1 &= \{ (U_1^n(M_1,L_1),U_2^n(M_2,L_2),Y_1^n) \notin \aepvar(U_1,U_2,Y_1) \}, \\
\Ec_2 &= \{ (U_1^n(m_1,l_1),U_2^n(m_2,l_2),Y_1^n) \in \aepvar(U_1,U_2,Y_1) \textrm{ for some } m_1 \neq M_1, \\ 
&\qquad\qquad\qquad\qquad\qquad \textrm{ for some } (m_2,l_1,l_2) \in [2^{nR_2}] \times [2^{n\Rh_1}] \times [2^{n\Rh_2}]\}.
\end{align*}
By the union of events bound, $\P_e^{(n)}(\Cc_n) \le \P(\Ec_0) + \P(\Ec_1 \cap \Ec_0^c) + \P(\Ec_2 \cap \Ec_0^c)$. Since $\Rh_1 + \Rh_2 = I(U_1;U_2) + 10 \e H(U_1,U_2)$, by the mutual covering lemma in~\cite[p.~208]{El-Gamal--Kim2011}, the probability $\P(\Ec_0)$ tends to zero as $n \to \infty$. By the conditional typicality lemma in~\cite[p.~27]{El-Gamal--Kim2011}, the probability $\P(\Ec_1 \cap \Ec_0^c)$ tends to zero as $n \to \infty$. The last term can be bounded by two ways. First, by the symmetric codebook generation,
\begin{align*}
\P(\Ec_{2} \cap \Ec_0^c) &\le \P(\Ec_2) \\
&=  \P(\Ec_2 | M_1=M_2=1) \\
&\le \P( (U_1^n(m_1,l_1),Y_1^n) \in \aepvar(U_1,Y_1) \textrm{ for some } m_1 \neq 1, \textrm{ for some } l_1 \in [2^{n\Rh_1}] | M_1 = 1),
\end{align*}
which tends to zero as $n \to \infty$ if $R_1 + \Rh_1 \le I(U_1;Y_1) - \d(\e')$ by the packing lemma in \cite{El-Gamal--Kim2011}. Letting $\Rh_1 = \a (I(U_1;U_2) + 10\e H(U_1,U_2))$, we have
\begin{equation}
R_1 \le \max\{0, I(U_1;Y_1) - \a I(U_1;U_2) - 2\d(\e')\}.
\label{eq:snd_1}
\end{equation}

Secondly, we can decompose the event $\Ec_2 = \Ec_{21} \cup \Ec_{22}$ such that
\begin{align*}
\Ec_{21} &= \{ (U_1^n(m_1,l_1),U_2^n(M_2,l_2),Y_1^n) \in \aepvar(U_1,U_2,Y_1) \textrm{ for some } m_1 \neq M_1, \\ 
&\qquad\qquad\qquad\qquad\qquad \textrm{ for some } (l_1,l_2) \in [2^{n\Rh_1}] \times [2^{n\Rh_2}]\},
\\
\Ec_{22} &= \{ (U_1^n(m_1,l_1),U_2^n(m_2,l_2),Y_1^n) \in \aepvar(U_1,U_2,Y_1) \textrm{ for some } m_1 \neq M_1,\textrm{ for some } m_2 \neq M_2, \\ 
&\qquad\qquad\qquad\qquad\qquad \textrm{ for some } (l_1,l_2) \in [2^{n\Rh_1}] \times [2^{n\Rh_2}]\}.
\end{align*}
We start with bounding $\P(\Ec_{22})$ as follows:
{\allowdisplaybreaks
\begin{align*}
\P(\Ec_{22}) &= \P(\Ec_{22}|M_1=M_2=1)
\\
&= \P((U_1^n(m_1,l_1),U_2^n(m_2,l_2),Y_1^n) \in \aepvar(U_1,U_2,Y_1) \textrm{ for some } m_1 \neq 1,\textrm{ for some } m_2 \neq 1, \\ 
&\qquad\qquad\qquad\qquad\qquad \textrm{ for some } (l_1,l_2) \in [2^{n\Rh_1}] \times [2^{n\Rh_2}] |M_1=M_2=1)
\\
&\overset{(a)}{\le} \sum_{m_1 \neq 1} \sum_{l_1} \sum_{m_2 \neq 1} \sum_{l_2} \sum_{(u_1^n,u_2^n,y_1^n) \atop \in \aepvar(U_1,U_2,Y_1)} p(y_1^n|M_1=M_2=1) 2^{-n(H(U_1)+H(U_2)-\d(\e'))}
\\
&\le \sum_{m_1 \neq 1} \sum_{l_1} \sum_{m_2 \neq 1} \sum_{l_2} 2^{-n(H(U_1)+H(U_2)-H(U_1,U_2|Y_1)-2\d(\e'))}
\\
&\le 2^{n(R_1+R_2+\Rh_1+\Rh_2)}  2^{-n(H(U_1)+H(U_2)-H(U_1,U_2|Y_1)-2\d(\e'))},
\end{align*}
}%
where $(a)$ follows since given $\{M_1=M_2=1\}$, the pair $(U_1^n(m_1,l_1),U_2^n(m_2,l_2))$ for $m_1 \neq 1, m_2 \neq 1$ is i.i.d. with respect to the product pmf $p(u_1)p(u_2)$ and is independent from $Y_1^n$. Substituting $\Rh_1 + \Rh_2 = I(U_1;U_2) + 10 \e H(U_1,U_2)$, it follows that $\P(\Ec_{22})$ tends to zero as $n \to \infty$ if $R_1 + R_2 \le I(U_1,U_2;Y_1)- 3\d(\e')$. 

We next bound the probability $\P(\Ec_{21} \cap \Ec_0^c)$. Define the events $\Mc_1 :=\{M_1=M_2=1\}$ and $\Mc_2 :=\{L_1=L_2=1\}$. By the symmetric codebook generation, 
\[
\P(\Ec_{21} \cap \Ec_0^c) = \P(\Ec_{21} \cap \Ec_0^c |\Mc_1,\Mc_2 ),
\]
which can be bounded as
\begin{align}\notag
&\P(\Ec_{21} \cap \Ec_0^c| \Mc_1,\Mc_2) \\ \notag
&\le \sum_{m_1 \neq 1} \sum_{l_1,l_2} \P( (U_1^n(m_1,l_1), U_2^n(1,l_2),Y_1^n) \in \aepvar, (U_1^n(1,1),U_2^n(1,1)) \in \aep | \Mc_1,\Mc_2) \\ \notag
&\le \sum_{m_1 \neq 1} \sum_{l_1} \P( (U_1^n(m_1,l_1), U_2^n(1,1),Y_1^n) \in \aepvar, (U_1^n(1,1),U_2^n(1,1)) \in \aep | \Mc_1,\Mc_2) + \\
&\qquad \sum_{m_1 \neq 1} \sum_{l_1} \sum_{l_2 \neq 1} \P( (U_1^n(m_1,l_1), U_2^n(1,l_2),Y_1^n) \in \aepvar, (U_1^n(1,1),U_2^n(1,1)) \in \aep | \Mc_1,\Mc_2).
\label{eq:sum12}
\end{align}
The first summation term in (\ref{eq:sum12}) can be bounded as
{\allowdisplaybreaks
\begin{align*}
&\sum_{m_1 \neq 1} \sum_{l_1} \P( (U_1^n(m_1,l_1), U_2^n(1,1),Y_1^n) \in \aepvar, (U_1^n(1,1),U_2^n(1,1)) \in \aep |\Mc_1,\Mc_2)
\\
&\le \sum_{m_1 \neq 1} \sum_{l_1} \sum_{(u_1^n,u_2^n) \atop \in \aep} \sum_{(\ut_1^n,y_1^n) \atop \in \aepvar(U_1,Y_1|u_2^n) } \P( U_1^n(m_1,l_1) = \ut_1^n, U_1^n(1,1)=u_1^n, U_2^n(1,1) = u_2^n,Y_1^n = y_1^n |\Mc_1,\Mc_2) 
\\
&\overset{(a)}{=} \sum_{m_1 \neq 1} \sum_{l_1} \sum_{(u_1^n,u_2^n) \atop \in \aep} \sum_{(\ut_1^n,y_1^n) \atop \in \aepvar(U_1,Y_1|u_2^n) } \P( U_1^n(m_1,l_1) = \ut_1^n, U_1^n(1,1)=u_1^n, U_2^n(1,1) = u_2^n|\Mc_1,\Mc_2) p(y_1^n|u_1^n,u_2^n)
\\
&\overset{(b)}{\le} 2^{n(\Rh_1+\Rh_2)} \sum_{m_1 \neq 1} \sum_{l_1} \sum_{(u_1^n,u_2^n) \atop \in \aepvar} \sum_{(\ut_1^n,y_1^n) \atop \in \aepvar(U_1,Y_1|u_2^n) } \P( U_1^n(m_1,l_1) = \ut_1^n, U_1^n(1,1)=u_1^n, U_2^n(1,1) = u_2^n|\Mc_1) p(y_1^n|u_1^n,u_2^n)
\\
&\overset{(c)}{\le} 2^{n(\Rh_1+\Rh_2)} \sum_{m_1 \neq 1} \sum_{l_1} \sum_{(u_1^n,u_2^n) \atop \in \aepvar} \sum_{(\ut_1^n,y_1^n) \atop \in \aepvar(U_1,Y_1|u_2^n) } p(y_1^n|u_1^n,u_2^n) 2^{-n(2H(U_1)+H(U_2) -\d(\e'))}
\\
&\le 2^{n(\Rh_1+\Rh_2)} \sum_{m_1 \neq 1} \sum_{l_1} \sum_{(u_1^n,u_2^n) \atop \in \aepvar} 2^{n(H(U_1|Y_1,U_2) + \d(\e')) }  2^{-n(2H(U_1)+H(U_2) -\d(\e'))}
\\
&\le 2^{n(\Rh_1+\Rh_2)} \sum_{m_1 \neq 1} \sum_{l_1} 2^{n(H(U_1,U_2)+\d(\e'))} 2^{n(H(U_1|Y_1,U_2) + \d(\e')) }  2^{-n(2H(U_1)+H(U_2) -\d(\e'))}
\\
&\le 2^{n(R_1+2\Rh_1+\Rh_2+H(U_1,U_2)+H(U_1|Y_1,U_2) -2H(U_1)-H(U_2) + 3\d(\e'))}
\\
&= 2^{n(R_1+2\Rh_1+\Rh_2-I(U_1;U_2)-I(U_1;Y_1,U_2) + 3\d(\e'))},
\end{align*}
}%
where $(a)$ follows since given $(\Mc_1,\Mc_2)$ the tuple $U_1^n(m_1,l_1) \to (U_1^n(1,1),U_2^n(1,1)) \to Y_1^n$ form a Markov chain, $(b)$ follows by~\cite[Lemma 11]{Lim--Gastpar2016}, and $(c)$ follows since the tuple $(U_1^n(m_1,l_1),U_1^n(1,1),U_2^n(1,1))$ is independent of the event $\Mc_1$ and is i.i.d. with respect to the product pmf $p(u_1)p(u_1)p(u_2)$.

Similarly, the second summation term in (\ref{eq:sum12}) can be bounded as
\begin{align*}
&\sum_{m_1 \neq 1} \sum_{l_1} \sum_{l_2 \neq 1} \P( (U_1^n(m_1,l_1), U_2^n(1,l_2),Y_1^n) \in \aepvar, (U_1^n(1,1),U_2^n(1,1)) \in \aep |\Mc_1,\Mc_2)
\\
&\le 2^{n(R_1+2\Rh_1+2\Rh_2-2I(U_1;U_2)-I(U_1,U_2;Y_1) + 3\d(\e'))}.
\end{align*}
Therefore, $\P(\Ec_{21} \cap \Ec_0^c)$ tends to zero as $n \to \infty$ if $R_1+2\Rh_1+\Rh_2 \le I(U_1;U_2)+I(U_1;Y_1,U_2) - 3\d(\e')$ and $R_1+2\Rh_1+2\Rh_2 \le 2I(U_1;U_2)+I(U_1,U_2;Y_1)-3\d(\e')$. Letting $\Rh_1 = \a (I(U_1;U_2) + 10\e H(U_1,U_2))$ and $\Rh_2 = \ab (I(U_1;U_2) + 10\e H(U_1,U_2))$ results in $R_1 \le I(U_1;Y_1,U_2) - \a I(U_1;U_2) - 4\d(\e')$ and $R_1 \le I(U_1,U_2;Y_1)-4\d(\e')$. 

Combining with (\ref{eq:snd_1}), the probability of error at Decoder 1 tends to zero as $n \to \infty$ if 
\begin{equation}
\label{eq:marton11}
R_1 \le \max\{0, I(U_1;Y_1) - \a I(U_1;U_2) - 4\d(\e')\},
\end{equation}
or
\begin{subequations}
\label{eq:marton12}
\begin{align}
R_1 &\le I(U_1;Y_1,U_2) - \a I(U_1;U_2) - 4\d(\e'),\\
R_1 +  R_2 &\le I(U_1,U_2;Y_1) - 4\d(\e').
\end{align}
\end{subequations}

Repeating similar steps, the probability of error at Decoder 2 tends to zero as $n \to \infty$ if 
\begin{equation}
\label{eq:marton21}
R_2 \le \max\{0, I(U_2;Y_2) - \ab I(U_1;U_2) - 4\d(\e')\},
\end{equation}
or
\begin{subequations}
\label{eq:marton22}
\begin{align}
R_2 &\le I(U_2;Y_2,U_1) - \ab I(U_1;U_2) - 4\d(\e'),\\
R_1 +  R_2 &\le I(U_1,U_2;Y_2) - 4\d(\e').
\end{align}
\end{subequations}

If we denote the set of rate pairs satisfying (\ref{eq:marton11}) or (\ref{eq:marton12}) as $\Rrbci(p, \a, \d(\e'))$, and denote the set of rate pairs satisfying (\ref{eq:marton21}) or (\ref{eq:marton22}) as $\Rrbcii(p, \a, \d(\e'))$, then the rate region $\Rrbci(p,\a,\d(\e')) \cap \Rrbcii(p,\a,\d(\e'))$ is achievable by the $\e'$-typicality decoders. Define the rate regions $\Rrbcj(p,\a) := \Rrbcj(p,\a,\d(\e')=0)$, $j=1,2$. Let $\e'=2\e$. Taking $\e \to 0$ and then taking the closure implies 
\[
\Rrbci(p,\a) \cap \Rrbcii(p,\a) \; \sbq \;  \Rrbc^*(p,\a).
\]
The achievability proof follows from the next lemma that provides an equivalent characterization for the rate region in Theorem \ref{thm:opt_marton}.
\begin{lemma}
\label{lem:equiv_reg_marton}
For any input pmf $p=p(u_1,u_2)$, function $x(u_1,u_2)$, and $\a \in [0 \; 1]$, 
\[
\Rrbc^{**}(p,\a) = \Rrbci(p,\a) \cap \Rrbcii(p,\a).
\]
\end{lemma}

\begin{IEEEproof}
Fix pmf $p = p(u_1,u_2)$, function $x(u_1,u_2)$ and $\a \in [0 \; 1]$. It suffices to show that the rate region $\Rrbci(p,\a)$ is equivalent to the set of rate pairs $(R_1,R_2)$ that satisfy (\ref{eq:thm_marton1})-(\ref{eq:thm_marton2}). We first show that any rate pair in $\Rrbci(p,\a)$ satisfies (\ref{eq:thm_marton1})-(\ref{eq:thm_marton2}). Suppose that the rate pair $(R_1, R_2) \in \Rrbci(p,\a)$, which implies that
\begin{align*}
R_1 \le I(U_1;Y_1,U_2) - \a I(U_1;U_2),
\end{align*}
and
\begin{align*}
R_1 &\le \max\{0, I(U_1;Y_1) - \a I(U_1;U_2), I(U_1,U_2;Y_1) - R_2\} \\
&= I(U_1,U_2;Y_1) - \min \{I(U_1,U_2;Y_1), I(U_2;Y_1,U_1) - \ab I(U_1;U_2), R_2\}.
\end{align*}
Therefore, $(R_1,R_2)$ satisfies (\ref{eq:thm_marton1})-(\ref{eq:thm_marton2}).

For the other direction, suppose that the rate pair $(R_1, R_2)$ satisfies (\ref{eq:thm_marton1})-(\ref{eq:thm_marton2}). Assume also that $R_2 < \min\{I(U_2;Y_1,U_1)-\ab I(U_1;U_2), I(U_1,U_2;Y_1)\}$. It then follows that
\begin{align*}
R_1 &\le I(U_1;Y_1,U_2) - \a I(U_1;U_2),\\
R_1 &\le I(U_1,U_2;Y_1) - R_2.
\end{align*}
So, $(R_1,R_2) \in \tRr(p,\a)$. If instead $R_2 \ge \min\{I(U_2;Y_1,U_1)-\ab I(U_1;U_2), I(U_1,U_2;Y_1)\}$, then
\begin{align*}
R_1 &\le I(U_1;Y_1,U_2) - \a I(U_1;U_2),\\
R_1 &\le I(U_1,U_2;Y_1) - \min\{I(U_2;Y_1,U_1)-\ab I(U_1;U_2), I(U_1,U_2;Y_1)\} = \max\{0, I(U_1;Y_1) - \a I(U_1;U_2) \}.
\end{align*}
Therefore, $(R_1,R_2) \in \Rrbci(p,\a)$, which completes the proof of the lemma.
\end{IEEEproof}

\section{Proof of Lemma~\ref{lem:outer2_list_marton}}
\label{app:list_marton}

Let $\e' > \e$. First, by (the averaged version of) Fano's lemma in (\ref{eq:marton_fano}), we have
\[
I(M_2; Y_1^n |\Cc_n) \ge I(M_2; M_1, Y_1^n | \Cc_n) - n \e_n.
\]
Therefore, it suffices to prove that for $n$ sufficiently large,
\[
I(M_2; M_1, Y_1^n |\Cc_n) \ge n [\min \{ R_2, I(U_2;Y_1,U_1) - \ab I(U_1;U_2), I(U_1,U_2;Y_1) \} - \d(\e') - 2\e_n],
\]
for some $\d(\e')$ that tends to zero as $\e \to 0$.

Similar to \cite{Bandemer--El-Gamal--Kim2012a}, we will show that given $M_1, Y_1^n$ and $\Cc_n$, a relatively short list $\Lc \sbq [2^{nR_2}]$ can be constructed that contains $M_2$ with high probability. Define a random set
\begin{align*}
\Lc &= \{ m_2 \in [2^{nR_2}]: (U_1^n(M_1,l_1),U_2^n(m_2,l_2), Y_1^n) \in \aepvar(U_1,U_2, Y_1) \\
&\qquad\qquad\qquad\qquad \textrm{ for some } (l_1,l_2) \in [2^{n\Rh_1}] \times [2^{n\Rh_2}]\}.
\end{align*}

Define the events $\Mc_1 = \{M_1 = M_2 = 1\}$ and $\Mc_2=\{L_1=L_2=1 \}$. The indicator random variable $\Et_n$ is as defined in (\ref{eq:indicator_en_marton}). By the symmetry of the codebook generation, for each $m_2 \neq M_2 \in [2^{nR_2}]$ we start with 
\begin{align} \notag
& \P( m_2 \in \Lc, \Et_n = 1) \\ \notag
&= \P( m_2 \in \Lc, \Et_n=1 | \Mc_1,\Mc_2) 
\\ \notag
& \overset{(a)}{=}  \P( (U_1^n(1,l_1), U_2^n(m_2,l_2), Y_1^n) \in \aepvar \textrm{ for some } (l_1,l_2) \in [2^{n\Rh_1}] \times [2^{n\Rh_2}], \\ \notag
&\qquad\qquad\qquad\qquad\qquad\qquad\qquad (U_1^n(1,1), U_2^n(1,1)) \in \aep  | \Mc_1,\Mc_2)\\  \notag
& \overset{(b)}{\le} \sum_{l_2} \sum_{(u_1^n,u_2^n) \in \atop \aep(U_1,U_2)} \sum_{(\ut_2^n,y_1^n) \in \atop \aepvar(U_2,Y_1|u_1^n)} 
\P ( U_{1}^n(1,1) = u_1^n, U_{2}^n(1,1) = u_2^n, U_{2}^n(m_2,l_2) = \ut_2^n, Y_1^n = y_1^n | \Mc_1,\Mc_2) 
\\ 
&\quad + \sum_{l_1\neq 1} \sum_{l_2} \sum_{(u_1^n,u_2^n) \in \atop \aep(U_1,U_2)} \sum_{(\ut_1^n,\ut_2^n,y_1^n) \in \atop \aepvar(U_1,U_2,Y_1)} 
\P \left(\begin{array}{c|c}
U_{1}^n(1,1) = u_1^n, U_{2}^n(1,1) = u_2^n, & \Mc_1, \\
U_{1}^n(m_1,l_1) = \ut_1^n, U_{2}^n(m_2,l_2) = \ut_2^n, Y_1^n = y_1^n &  \Mc_2
\end{array}\right)
\label{eq:marton_two_sum}
\end{align}
where $(b)$ follows by the union of events bound and by decomposing the event in $(a)$ onto two sets: $\{l_1=1\}$ and $\{l_1 \neq 1\}$. Two summation terms on the right hand side of (\ref{eq:marton_two_sum}) can be bounded using techniques similar to those in the achievability proof (see Appendix~\ref{app:marton_achiev}) for Theorem \ref{thm:opt_marton} to get
\[
\P( m_2 \in \Lc, \Et_n = 1) \le 2^{-n(I(U_2;Y_1,U_1)-\ab I(U_1;U_2) - 4\d(\e'))} + 2^{-n(I(U_1,U_2;Y_1)-4\d(\e'))}.
\] 

Since $\P(\Et_n = 1)$ tends to one as $n \to \infty$, for $n$ sufficiently large, $ \P( m_2 \in \Lc | \Et_n = 1) \le \P( m_2 \in \Lc , \Et_n = 1) q^{\e}$. The expected cardinality of $\Lc$ given  $\{\Et_n = 1 \}$ is then bounded as
\begin{align} \notag
\E(|\Lc| | \Et_n = 1) & \le 1 + \sum_{m_2 \neq M_2} \P( m_2 \in \Lc | \Et_n = 1) 
\\ \notag
& \le 1+ 2^{n(R_2 - I(U_2;Y_1,U_1)+\ab I(U_1;U_2) + 4\d(\e')+\frac{\e}{n})} + 2^{n(R_2-I(U_1,U_2;Y_1)+4\d(\e') + \frac{\e}{n})}
\\
&= 1+ 2^{n(R_2 - I(U_2;Y_1,U_1)+\ab I(U_1;U_2) + 4\d(\e')+\e_n)} + 2^{n(R_2-I(U_1,U_2;Y_1)+4\d(\e') + \e_n)}
\label{eq:ent_list_marton}
\end{align}
for $n$ sufficiently large. 

Define another indicator random variable $\Ft_n = \mathbbm{1}_{\{ M_2 \in \Lc \}}$. Since $\e' > \e$ and $\P(\Et_n=1)$ tends to one as $n \to \infty$, by the conditional typicality lemma in~\cite[p.~27]{El-Gamal--Kim2011}, $\P(\Ft_n=1)$ tends to one as $n \to \infty$. Then, for $n$ sufficiently large, we have
\begin{align*} 
H(M_2 | \Cc_n, M_1, Y_1^n) 
&= H(M_2 | \Cc_n, M_1, Y_1^n, \Et_n,\Ft_n) + I ( M_2; \Et_n,\Ft_n | \Cc_n, M_1, Y_1^n)
\\ 
& \le H(M_2 | \Cc_n, M_1, Y_1^n, \Et_n,\Ft_n) + 2
\\ 
& \le 2 + \P(\Ft_n=0) H(M_2 | \Cc_n, M_1, Y_1^n, \Et_n,\Ft_n=0) + H(M_2 | \Cc_n, M_1, Y_1^n, \Et_n,\Ft_n=1)
\\
& \le 2 +  nR_2 \P(\Ft_n=0) + H(M_2 | \Cc_n, M_1^n, Y_1^n, \Et_n,\Ft_n=1).
\end{align*}

For the last term, we use the fact that if $M_2 \in \Lc$, then the conditional entropy cannot exceed $\log(|\Lc|)$:
{\allowdisplaybreaks
\begin{align*}
&H(M_2 |  \Cc_n, M_1, Y_1^n, \Et_n,\Ft_n=1) 
\\
& \overset{(a)}{=} H(M_2 | \Cc_n, M_1, Y_1^n, \Et_n,\Ft_n=1, \Lc, |\Lc|) 
\\
& \le H(M_2 | \Et_n,\Ft_n=1, \Lc, |\Lc|) 
\\
&= \sum_{l=0}^{2^{nR_2}} \P(|\Lc|=l,\Et_n=1) H(M_2 | \Et_n=1, \Ft_n = 1, \Lc, |\Lc| = l) \\
&\quad\quad\quad + \sum_{l=0}^{2^{nR_2}} \P(|\Lc|=l,\Et_n=0) H(M_2 | \Et_n=0, \Ft_n = 1, \Lc, |\Lc| = l)
\\
&\le \sum_{l=0}^{2^{nR_2}} \P(|\Lc|=l,\Et_n=1) H(M_2 | \Et_n=1, \Ft_n = 1, \Lc, |\Lc| = l) +  {nR_2} \P(\Et_n=0)
\\
&\le \sum_{l=0}^{2^{nR_2}} \P(|\Lc|=l,\Et_n=1) \log(l) +  {nR_2} \P(\Et_n=0)
\\
&\le \sum_{l=0}^{2^{nR_2}} \P(|\Lc|=l|\Et_n=1) \log(l) +  {nR_2} \P(\Et_n=0)
\\
&= \E [\log(|\Lc|) | \Et_n=1] +  {nR_2} \P(\Et_n=0)
\\
& \overset{(b)}{\le} \log(\E [|\Lc| | \Et_n=1]) +  {nR_2} \P(\Et_n=0) 
\\
& \overset{(c)}{\le}  \max \{0, n(R_2 - I(U_2;Y_1,U_1)+\ab I(U_1;U_2)+ 4\d(\e')+\e_n), n(R_2 - I(U_1,U_2;Y_1) + 4\d(\e')+\e_n) \} \\
&\qquad\qquad\qquad + {nR_2} \P(\Et_n=0)
\\
& \le n \cdot \max \{0, R_2 - I(U_2;Y_1,U_1)+\ab I(U_1;U_2), R_2 - I(U_1,U_2;Y_1) \} + n 4\d(\e') + n\e_n + {nR_2} \P(\Et_n=0),
\end{align*}
}%
where $(a)$ follows since the set $\Lc$ and its cardinality $| \Lc|$ are functions of $(\Cc_n, M_1, Y_1^n)$, $(b)$ follows by Jensen's inequality, and $(c)$ follows by (\ref{eq:ent_list_marton}) and the soft-max interpretation of the log-sum-exp function~\cite[p.~72]{Boyd--Vandenberghe2004}.
Substituting back gives
\begin{align*}
 I(M_2; M_1, Y_1^n | \Cc_n) 
&= H(M_2 | \Cc_n) - H(M_2 | \Cc_n, M_1, Y_1^n)
\\
&= nR_2 - H(M_2 | \Cc_n, M_1, Y_1^n)
\\
& \ge nR_2 - 2 - nR_2 \P(\Ft_n=0) - H(M_2 | \Cc_n, M_1^n, Y_1^n, \Et_n,\Ft_n=1)
\\
& \ge nR_2 - 2 - nR_2 \P(\Ft_n=0) -  n 4\d(\e') - n\e_n - {nR_2} \P(\Et_n=0) \\
&\quad\quad
-n \cdot \max \{0, R_2 - I(U_2;Y_1,U_1)+\ab I(U_1;U_2), R_2 - I(U_1,U_2;Y_1) \} 
\\
&\overset{(a)}{=} n [ \min \{ R_2, I(U_2;Y_1,U_1)-\ab I(U_1;U_2), I(U_1,U_2;Y_1)\} - 4\d(\e') - 2\e_n ],
\end{align*}
where $(a)$ follows since both of the probabilities $\P(\Et_n=0)$ and $\P(\Ft_n=0)$ tend to zero as $n \to \infty$.

\bibliography{nit}
\end{document}